\documentclass[10pt,journal,compsoc]{IEEEtran}
\ifCLASSOPTIONcompsoc
  \usepackage[nocompress]{cite}

\else
  \usepackage{cite}
\fi
\ifCLASSINFOpdf
   \usepackage[pdftex]{graphicx}
\else
\fi
\usepackage{url}

\usepackage{hyperref}  
\hypersetup{
    hidelinks = true
}

\usepackage{amsbsy}

\usepackage{amssymb}

\usepackage{enumitem}

\usepackage{csquotes}

\usepackage[dvipsnames]{xcolor}

\usepackage{rotating}
\usepackage{colortbl}
\usepackage{tcolorbox}
\usepackage{tikz}

\newcommand{\todo}[1]{}
\newcommand{\review}[1]{}
\newcommand{\fred}[1]{}
\newcommand{\Brian}[1]{}
\newcommand{\robert}[1]{}
\newcommand{\polo}[1]{}

\newcommand{\hide}[1]{}

\usepackage{array}
\newcolumntype{L}[1]{>{\raggedright\let\newline\\\arraybackslash\hspace{0pt}}m{#1}}
\newcolumntype{C}[1]{>{\centering\let\newline\\\arraybackslash\hspace{0pt}}m{#1}}
\newcolumntype{R}[1]{>{\raggedleft\let\newline\\\arraybackslash\hspace{0pt}}m{#1}}
\newcolumntype{P}[1]{>{\raggedright}p{#1}}

\usepackage{booktabs}

\renewcommand{\arraystretch}{1.8}

\definecolor{why}{HTML}{EE220C}
\definecolor{who}{HTML}{00A2FF}
\definecolor{what}{HTML}{1DB100}
\definecolor{how}{HTML}{F8BA00}
\definecolor{when}{HTML}{CB297B}
\definecolor{where}{HTML}{00A89D}

\definecolor{white}{HTML}{FFFFFF}
\definecolor{black}{HTML}{000000}
\definecolor{tabletagcolor}{HTML}{BDBDBD}

\definecolor{cell}{HTML}{B0BEC5}
\definecolor{rowbackground}{HTML}{F0F2F4}
\newcommand{\f}{
\begin{tikzpicture}[every node/.style={inner sep=0,outer sep=0},scale=0.4]
    \fill [rounded corners=0.08cm,fill=cell] (0,0)--(.6,0)--(.6,.6)--(0,.6)--cycle;
\end{tikzpicture}
}

\newcommand{\rspace}{$\hspace{0.1cm}$}

\usepackage{tcolorbox}

\definecolor{tagbordercolor}{rgb}{0.8, 0.8, 0.8}
\definecolor{tagbgcolor}{rgb}{0.9, 0.9, 0.9}

\newtcbox{\whytag}{nobeforeafter, colframe=why,
colback=why, boxrule=0.5pt, arc=1pt,
 boxsep=0pt,left=2pt,right=2pt,top=1.5pt,bottom=2pt,tcbox raise base}

\newtcbox{\whotag}{nobeforeafter, colframe=who,
colback=who, boxrule=0.5pt, arc=1pt,
 boxsep=0pt,left=2pt,right=2pt,top=1.5pt,bottom=2pt,tcbox raise base}

\newtcbox{\whattag}{nobeforeafter, colframe=what,
colback=what, boxrule=0.5pt, arc=1pt,
 boxsep=0pt,left=2pt,right=2pt,top=1.5pt,bottom=2pt,tcbox raise base}

\newtcbox{\howtag}{nobeforeafter, colframe=how,
colback=how, boxrule=0.5pt, arc=1pt,
 boxsep=0pt,left=2pt,right=2pt,top=1.5pt,bottom=2pt,tcbox raise base}

\newtcbox{\whentag}{nobeforeafter, colframe=when,
colback=when, boxrule=0.5pt, arc=1pt,
 boxsep=0pt,left=2pt,right=2pt,top=1.5pt,bottom=2pt,tcbox raise base}

\newtcbox{\wheretag}{nobeforeafter, colframe=where,
colback=where, boxrule=0.5pt, arc=1pt,
 boxsep=0pt,left=2pt,right=2pt,top=1.5pt,bottom=2pt,tcbox raise base}

\newtcbox{\tabletag}{nobeforeafter, colframe=tabletagcolor,
colback=white, boxrule=0.5pt, arc=1pt,
 boxsep=0pt,left=2pt,right=2pt,top=1.5pt,bottom=2pt,tcbox raise base}

\begin{document}
%
\title{Visual Analytics in Deep Learning:\\An Interrogative Survey for the Next Frontiers}
%
%
%
%

\author{Fred~Hohman,~\IEEEmembership{Member,~IEEE,}
        Minsuk~Kahng,~\IEEEmembership{Member,~IEEE,}
        Robert~Pienta,~\IEEEmembership{Member,~IEEE,}
        and~Duen~Horng~Chau,~\IEEEmembership{Member,~IEEE}
\IEEEcompsocitemizethanks{\IEEEcompsocthanksitem F. Hohman, M. Kahng, R. Pienta, and D. H. Chau are with the College of Computing, Georgia Tech, Atlanta,
Georgia 30332, U.S.A.\protect\\
E-mail: \{fredhohman, kahng, pientars, polo\}@gatech.edu
}
}

%
%

\markboth{}%
{Hohman et al.: Visual Analytics in Deep Learning: An Interrogative Survey for the Next Frontiers}
%



\IEEEtitleabstractindextext{%
\begin{abstract}
Deep learning has recently seen rapid development and received significant attention due to its state-of-the-art performance on previously-thought hard problems.
However, because of the internal complexity and nonlinear structure of deep neural networks, the underlying decision making processes for why these models are achieving such performance are challenging and sometimes mystifying to interpret.
As deep learning spreads across domains, it is of paramount importance that we equip users of deep learning with tools for understanding when a model works correctly, when it fails, and ultimately how to improve its performance.
Standardized toolkits for building neural networks have helped democratize deep learning; visual analytics systems have now been developed to support model explanation, interpretation, debugging, and improvement.
We present a survey of the role of visual analytics in deep learning research, which highlights its short yet impactful history and thoroughly summarizes the state-of-the-art using a human-centered interrogative framework,
focusing on the \textit{Five W's and How} (Why, Who, What, How, When, and Where).
We conclude by highlighting research directions and open research problems.
This survey helps researchers and practitioners in both visual analytics and deep learning to quickly learn key aspects of this young and rapidly growing body of research, whose impact spans a diverse range of domains.
\end{abstract}

}

\maketitle

\IEEEdisplaynontitleabstractindextext

%
\IEEEpeerreviewmaketitle

\begin{figure*}[!h]
 \centering
 \includegraphics[width=0.97\textwidth]{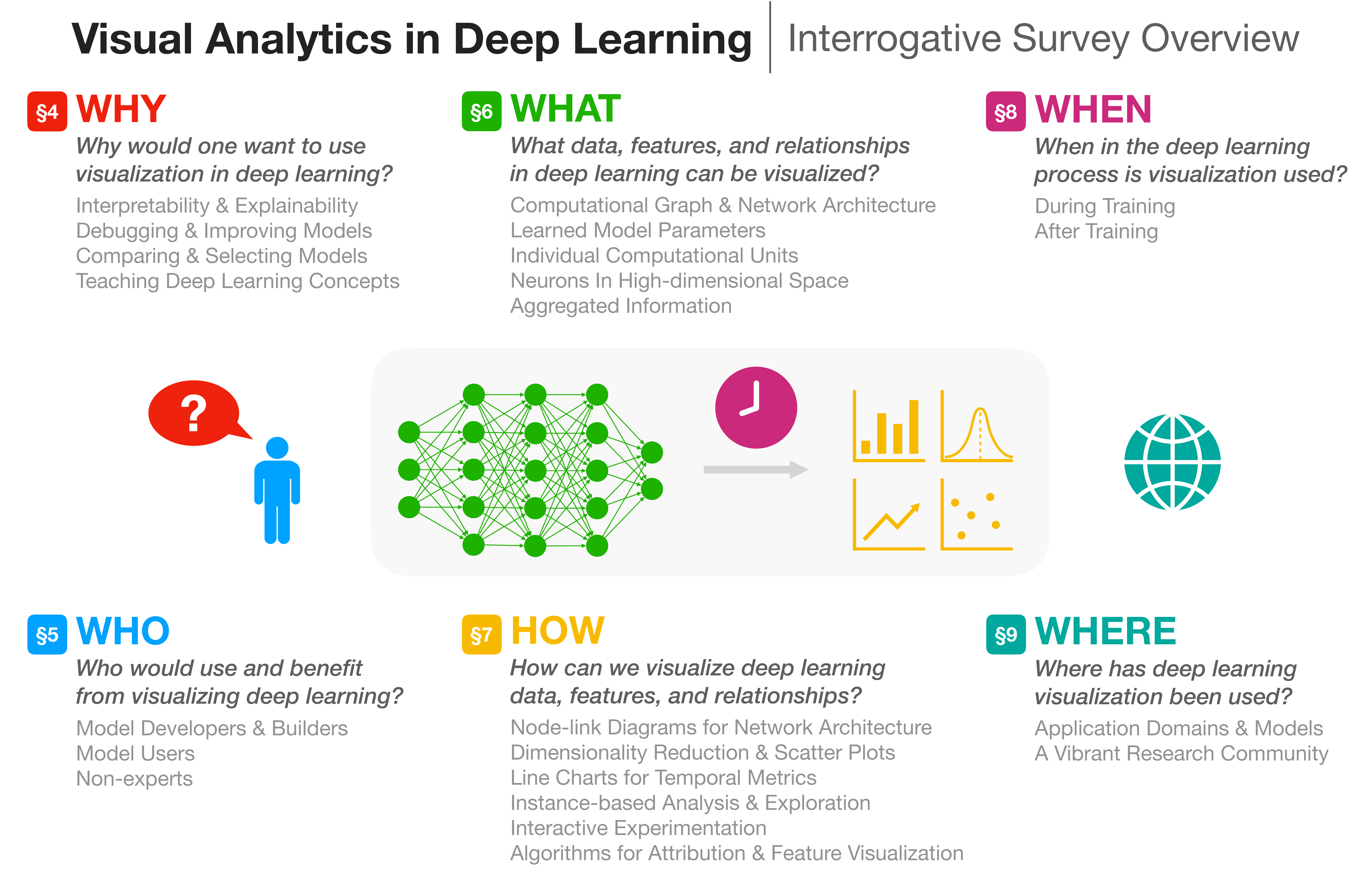}
 \caption{
 A visual overview of our interrogative survey, and how each of the six questions, "Why, Who, What, How, When, and Where," relate to one another.
 Each question corresponds to one section of this survey, indicated by the numbered tag, near each question title.
 Each section lists its major subsections discussed in the survey.
 }
 \label{fig:overview}
\end{figure*}

\IEEEraisesectionheading{\section{Introduction}
\label{sec:intro}}

%
%
%
%
\IEEEPARstart{D}{eep}
learning is a specific set of techniques from the broader field of machine learning (ML) that focus on the study and usage of \textit{deep} artificial neural networks to learn structured representations of data.
First mentioned as early as the 1940s~\cite{mcculloch1943logical}, artificial neural networks have a rich history~\cite{rawat2017deep}, and have recently seen a dominate and pervasive resurgence~\cite{krizhevsky2012imagenet, simonyan2013deep, szegedy2015going} in many research domains by producing state-of-the-art results~\cite{karpathy2014learned, he2016deep} on a number of diverse big data tasks~\cite{deng2009imagenet, russakovsky2015imagenet}.
For example, the premiere machine learning, deep learning, and artificial intelligence (AI) conferences have seen enormous growth in attendance and paper submissions since early 2010s.
Furthermore, open-source toolkits and programming libraries for building, training, and evaluating deep neural networks have become more robust and easy to use, democratizing deep learning.
As a result, the barrier to developing deep learning models is lower than ever before and deep learning applications are becoming pervasive.

While this technological progress is impressive, it comes with unique and novel challenges.
For example, the lack of interpretability and transparency of neural networks, from the learned representations to the underlying decision process, is an important problem to address.
Making sense of why a particular model misclassifies test data instances or behaves poorly at times is a challenging task for model developers.
Similarly, end-users interacting with an application that relies on deep learning to make critical decisions may question its reliability if no explanation is given by the model, or become baffled if the explanation is convoluted.
While explaining neural network decisions is important, there are numerous other problems that arise from deep learning, such as AI safety and security (e.g., when using models in applications such as self-driving vehicles), 
and compromised trust due to bias in models and datasets, just to name a few. 
These challenges are often compounded, due to the large datasets required to train most deep learning models.
As worrisome as these problems are, they will likely become even more widespread as more AI-powered systems are deployed in the world.
Therefore, a general sense of model understanding is not only beneficial, but often required to address the aforementioned issues.

Data visualization and visual analytics excel at knowledge communication and insight discovery by using encodings to transform abstract data into meaningful representations.
In the seminal work by Zeiler and Fergus~\cite{zeiler2014visualizing}, a technique called \textit{deconvolutional networks} enabled projection from a model's learned feature space back to the pixel space.
Their technique and results give insight into what types of features deep neural networks are learning at specific layers, and also serve as a debugging tool for improving a model.
This work is often credited for popularizing visualization in the machine learning and computer vision communities in recent years, putting a spotlight on it as a powerful tool that helps people understand and improve deep learning models.
However, visualization research for neural networks started well before~\cite{craven1992visualizing, streeter2001nvis, tzeng2005opening}.
Over just a handful of years, many different techniques have been introduced to help interpret what neural networks are learning.
Many such techniques generate static images, such as attention maps and heatmaps for image classification, indicating which parts of an image are most important to the classification.
However, interaction has also been incorporated into the model understanding process in visual analytics tools to help people gain insight~\cite{liu2017towards, wongsuphasawat2018visualizing, smilkov2017direct}.
This hybrid research area has grown in both academia and industry, forming the basis for many new research papers, academic workshops, and deployed industry tools.

In this survey, we summarize a large number of deep learning visualization works using the \textit{Five W's and How} (Why, Who, What, How, When, and Where).
Figure \ref{fig:overview} presents a visual overview of how these interrogative questions reveal and organize the various facets of deep learning visualization research and their related topics.
By framing the survey in this way, many existing works fit a description as the following fictional example:

\begin{displayquote}
\vspace{0.1cm}
\textit{To interpret representations learned by deep models (\textbf{\textcolor{why}{why}}), model developers (\textbf{\textcolor{who}{who}}) visualize neuron activations in convolutional neural networks (\textbf{\textcolor{what}{what}}) using t-SNE embeddings (\textbf{\textcolor{how}{how}}) after the training phase (\textbf{\textcolor{when}{when}}) to solve an urban planning problem (\textbf{\textcolor{where}{where}}).}
\vspace{0.1cm}
\end{displayquote}

\noindent
This framing captures the needs, audience, and techniques of deep learning visualization, and positions new work's contributions in the context of existing literature.

We conclude by highlighting prominent research directions and open problems. 
We hope that this survey acts as a companion text for researchers and practitioners wishing to understand how visualization supports deep learning research and applications.

\section{Our Contributions \& Method of Survey}
\label{sec:contributions}

\subsection{Our Contributions}

\begin{itemize}[itemsep=0.1cm]
\item[\textbf{C1.}] We present a comprehensive, timely survey on visualization and visual analytics in deep learning research, using a human-centered, interrogative framework.
This method enables us to position each work with respect to its  \textit{Five Ws and How} (Why, Who, What, How, When, and Where), and flexibly discuss and highlight existing works' multifaceted contributions.

\begin{itemize}[leftmargin=0.5cm, label=$\bullet$, itemsep=0.1cm]

    \item Our human-centered approach using the \textit{Five W's and How} --- based on how we familiarize ourselves with new topics in everyday settings --- enables readers to quickly grasp important facets of this young and rapidly growing body of research.
    
    \item Our interrogative process provides a framework to describe existing works, as well as a model to base new work off of.

\end{itemize}
    
\item[\textbf{C2.}] 
To highlight and align the cross-cutting impact that visual analytics has had on deep learning across a broad range of domains, 
our survey goes beyond visualization-focused venues,
extending a wide scope that encompasses most relevant works from many top venues in artificial intelligence, machine learning, deep learning, and computer vision.
We highlight how visual analytics has been an integral component in solving some of AI's biggest modern problems, such as neural network interpretability, trust, and security.

\item[\textbf{C3.}]
As deep learning, and more generally AI, touches more aspects of our daily lives, we highlight important research directions and open problems that we distilled from the survey.
These include improving the capabilities of visual analytics systems for furthering interpretability, conducting more effective design studies for evaluating system usability and utility, advocating humans' important roles in AI-powered systems, and promoting proper and ethical use of AI applications to benefit society.

\end{itemize}

\subsection{Survey Methodology \& Summarization Process}
We selected existing works from top computer science journals and conferences in visualization (e.g., IEEE Transactions on Visualization and Computer Graphics (TVCG)), visual analytics (e.g., IEEE Conference on Visual Analytics Science and Technology (VAST)) and deep learning (e.g., Conference on Neural Information Processing Systems (NIPS) and the International Conference on Machine Learning (ICML)).
Since deep learning visualization is relatively new, much of the relevant work has appeared in workshops at the previously mentioned venues; therefore, we also include those works in our survey.
Table \ref{table:venues} lists some of the most prominent publication venues and their acronyms.
We also inspected preprints posted on arXiv (\url{https://arxiv.org/}), an open-access, electronic repository of manuscript preprints, whose computer science subject has become a hub for new deep learning research.
Finally, aside from the traditional aforementioned venues, we include non-academic venues with significant attention such as Distill, industry lab research blogs, and research blogs of influential figures.
Because of the rapid growth of deep learning research and the lack of a perfect fit for publishing and disseminating work in this hybrid area, therefore, the inclusion of these non-traditional sources are important to review, as they are highly influential and impactful to the field.

\begin{table}[]
\sffamily

\centering
\caption{
Relevant visualization and AI venues, in the order of: journals, conferences, workshops, open access journals, and preprint repositories.
In each category, visualization venues precede AI venues.
}
\label{table:venues}

\setlength{\tabcolsep}{0pt}
\renewcommand\arraystretch{1.6}

\begin{tabular}{L{1cm}@{\hskip 0.1cm}L{7.8cm}}

\toprule
TVCG & IEEE Transactions on Visualization and Computer Graphics \\

VAST & IEEE Conference on Visual Analytics Science and Technology \\

InfoVis & IEEE Information Visualization \\

VIS & IEEE Visualization Conference (VAST+InfoVis+SciVis) \\

CHI & ACM Conference on Human Factors in Computing Systems \\

NIPS & Conference on Neural Information Processing Systems \\

ICML & International Conference on Machine Learning \\

CVPR & Conference on Computer Vision and Pattern Recognition \\

ICLR & International Conference on Learning Representations \\

VADL & IEEE VIS Workshop on Visual Analytics for Deep Learning \\

HCML  & CHI Workshop on Human Centered Machine Learning \\

IDEA & KDD Workshop on Interactive Data Exploration \& Analytics \\

& ICML Workshop on Visualization for Deep Learning \\

WHI & ICML Workshop on Human Interpretability in ML \\

& NIPS Workshop on Interpreting, Explaining and Visualizing Deep Learning \\

& NIPS Interpretable ML Symposium \\

FILM & NIPS Workshop on Future of Interactive Learning Machines \\

& ACCV Workshop on Interpretation and Visualization of Deep Neural Nets \\

& ICANN Workshop on Machine Learning and Interpretability \\

Distill & Distill: Journal for Supporting Clarity in Machine Learning \\

arXiv & arXiv.org e-Print Archive\\

\bottomrule

\end{tabular}
\end{table}

Visualization takes many forms throughout the deep learning literature.
This survey focuses on visual analytics for deep learning.
We also include related works from the AI and computer vision communities that contribute novel static visualizations.
So far, the majority of work surrounds convolutional neural networks (CNNs) and image data; 
more recent work has begun to visualize other models, e.g., recurrent neural
networks (RNNs), long short-term memory units (LSTMs), and generative adversarial networks (GANs).
For each work, we recorded the following information if present:
\begin{itemize}
    \item Metadata (title, authors, venue, and year published)
    \item General approach and short summary
    \item Explicit contributions
    \item Future work
    \item Design component (e.g. user-centered design methodologies, interviews, evaluation)
    \item Industry involvement and open-source code
\end{itemize}

With this information, we used the \textit{Five W's and How} (Why, Who, What, How, When, and Where) to organize these existing works and the current state-of-the-art of visualization and visual analytics in deep learning.

\subsection{Related Surveys}
While there is a larger literature for visualization for machine learning, including predictive visual analytics~\cite{lu2017recent, lu2017state, ren2017squares} and human-in-the-loop interactive machine learning~\cite{amershi2014power, sacha2016human}, to our knowledge there is no comprehensive survey of visualization and visual analytics for deep learning. 
Regarding deep neural networks, related surveys include a recent book chapter that
discusses visualization of deep neural networks related to the field of computer vision~\cite{seifert2017visualizations}, 
an unpublished essay that proposes a preliminary taxonomy for visualization techniques~\cite{zeng2017towards}, 
and an article that focuses on describing interactive model analysis, which mentions deep learning in a few contexts while describing a high-level framework for general machine learning models~\cite{liu2017towardsva}. 
A recent overview article by Choo and Liu~\cite{choo2018visual} is the closest in spirit to our survey.
Our survey provides wider coverage and more detailed analysis of the literature.

Different from all the related articles mentioned above, our survey provides a comprehensive, human-centered, and interrogative framework to describe deep learning visual analytics tools, discusses the new, rapidly growing community at large, and presents the major research trajectories synthesized from existing literature.

\subsection{Survey Overview \& Organization}
\label{sec:overview}

Section \ref{sec:background} introduces common deep learning terminology.
Figure \ref{fig:overview} shows a visual overview of this survey's structure and
Table \ref{table:papers} summarizes representative works.
Each interrogative question (Why, Who, What, How, When, and Where) is given its own section for discussion, ordered to best motivate why visualization and visual analytics in deep learning is such a rich and exciting area of research.

\begin{itemize}[leftmargin=0.87cm, itemsep=0.1cm, topsep=0.2cm]

    \item[
    \whytag{ \textbf{\textcolor{white}{$\pmb{\S}$ \ref{sec:why}}} }
    ] \textbf{Why do we want to visualize deep learning?}\\
    Why and for what purpose would one want to use visualization in deep learning?
    
    \item[
    \whotag{ \textbf{\textcolor{white}{$\pmb{\S}$ \ref{sec:who}}} }
    ] \textbf{Who wants to visualize deep learning?}\\
    Who are the types of people and users that would use and stand to benefit from visualizing deep learning?
    
    \item[
    \whattag{ \textbf{\textcolor{white}{$\pmb{\S}$ \ref{sec:what}}} }
    ] \textbf{What can we visualize in deep learning?}\\
    What data, features, and relationships are inherent to deep learning that can be visualized?
   
    \item[
    \howtag{ \textbf{\textcolor{white}{$\pmb{\S}$ \ref{sec:how}}} }
    ] \textbf{How can we visualize deep learning?}\\
    How can we visualize the aforementioned data, features, and relationships?
    
    \item[
    \whentag{ \textbf{\textcolor{white}{$\pmb{\S}$ \ref{sec:when}}} }
    ] \textbf{When can we visualize deep learning?}\\
    When in the deep learning process is visualization used and best suited?
    
    \item[
    \wheretag{ \textbf{\textcolor{white}{$\pmb{\S}$ \ref{sec:where}}} }
    ] \textbf{Where is deep learning visualization being used?}\\
    Where has deep learning visualization been used?

\end{itemize}

Section \ref{sec:open-challenges} presents research directions and open problems that we gathered and distilled from the literature survey. 
Section \ref{sec:conclusion} concludes the survey.

\section{Common Terminology}
\label{sec:background}

To enhance readability of this survey, 
and to provide quick references for readers new to deep learning, we have tabulated a sample of relevant and common deep learning terminology used in this work, shown in Table \ref{table:terminology-foundational}.
The reader may want to refer to Table \ref{table:terminology-foundational} throughout this survey for technical terms, meanings, and synonyms used in various contexts of discussion.
The table serves as an introduction and summarization of the state-of-the-art. 
For definitive technical and mathematical descriptions, 
we encourage the reader to refer to excellent texts on 
deep learning and neural network design, such as the \textit{Deep Learning} textbook~\cite{Goodfellow-et-al-2016}.

\begin{table*}[]
\sffamily

\centering

\caption{
Overview of representative works in visual analytics for deep learning.
Each row is one work; works are sorted alphabetically by first author's last name.
Each column corresponds to a subsection from the six interrogative questions.
A work's relevant subsection is indicated by a colored cell.
}

\label{table:papers}


\setlength{\tabcolsep}{0pt}
\renewcommand\arraystretch{1.30}

\begin{tabular}{
R{4.7cm}|
C{0.55cm}C{0.55cm}C{0.55cm}C{0.55cm}|
C{0.55cm}C{0.55cm}C{0.55cm}|
C{0.55cm}C{0.55cm}C{0.55cm}C{0.55cm}C{0.55cm}|
C{0.55cm}C{0.55cm}C{0.55cm}C{0.55cm}C{0.55cm}C{0.55cm}|
C{0.55cm}C{0.55cm}|
L{2.0cm}
}

\multicolumn{1}{c}{} &
\multicolumn{4}{c}{{\normalsize{\textbf{\textcolor{why}{WHY}}}}} &
\multicolumn{3}{c}{\normalsize{\textbf{\textcolor{who}{WHO}}}} &
\multicolumn{5}{c}{\normalsize{\textbf{\textcolor{what}{WHAT}}}} &
\multicolumn{6}{c}{\normalsize{\textbf{\textcolor{how}{HOW}}}} &
\multicolumn{2}{c}{\normalsize{\textbf{\textcolor{when}{WHEN}}}} &
\multicolumn{1}{c}{\normalsize{\textbf{\textcolor{where}{WHERE}}}}
\\

\multicolumn{1}{r}{\textbf{Work} \rspace} & 
\multicolumn{1}{c}{\rotatebox{90}{\tabletag{ \textbf{\ref{subsec:why-interpretability}} } Interpretability \& Explainability}} &
\multicolumn{1}{c}{\rotatebox{90}{\tabletag{ \textbf{\ref{subsec:why-debugging}} } Debugging \& Improving Models}} &
\multicolumn{1}{c}{\rotatebox{90}{\tabletag{ \textbf{\ref{subsec:why-comparing}} } Comparing \& Selecting Models}} &
\multicolumn{1}{c|}{\rotatebox{90}{\tabletag{ \textbf{\ref{subsec:why-teaching}} } Teaching Deep Learning Concepts}} &
\multicolumn{1}{c}{\rotatebox{90}{\tabletag{ \textbf{\ref{subsec:who-developer}} } Model Developers \& Builders}} &
\multicolumn{1}{c}{\rotatebox{90}{\tabletag{ \textbf{\ref{subsec:who-user}} } Model Users}} &
\multicolumn{1}{c|}{\rotatebox{90}{\tabletag{ \textbf{\ref{subsec:who-nonexpert}} } Non-experts}} &
\multicolumn{1}{c}{\rotatebox{90}{\tabletag{ \textbf{\ref{subsec:what-architecture}} } Computational Graph \& Network Architecture}} &
\multicolumn{1}{c}{\rotatebox{90}{\tabletag{ \textbf{\ref{subsec:what-parameters}} } Learned Model Parameters}} &
\multicolumn{1}{c}{\rotatebox{90}{\tabletag{ \textbf{\ref{subsec:what-individual}} } Individual Computational Units}} &
\multicolumn{1}{c}{\rotatebox{90}{\tabletag{ \textbf{\ref{subsec:what-highdimensional}} } Neurons in High-dimensional Space}} &
\multicolumn{1}{c|}{\rotatebox{90}{\tabletag{ \textbf{\ref{subsec:what-aggregated}} } Aggregated Information}} &
\multicolumn{1}{c}{\rotatebox{90}{\tabletag{ \textbf{\ref{subsec:how-architecture}} } Node-link Diagrams for Network Architecture}} &
\multicolumn{1}{c}{\rotatebox{90}{\tabletag{ \textbf{\ref{subsec:how-scatter}} } Dimensionality Reduction \& Scatter Plots}} &
\multicolumn{1}{c}{\rotatebox{90}{\tabletag{ \textbf{\ref{subsec:how-line}} } Line Charts for Temporal Metrics}} &
\multicolumn{1}{c}{\rotatebox{90}{\tabletag{ \textbf{\ref{subsec:how-instance}} } Instance-based Analysis \& Exploration}} &
\multicolumn{1}{c}{\rotatebox{90}{\tabletag{ \textbf{\ref{subsec:how-ie}} } Interactive Experimentation}} &
\multicolumn{1}{c|}{\rotatebox{90}{\tabletag{ \textbf{\ref{subsec:how-algorithms}} } Algorithms for Attribution \& Feature Visualization}} &
\multicolumn{1}{c}{\rotatebox{90}{\tabletag{ \textbf{\ref{subsec:when-during}} } During Training}} &
\multicolumn{1}{c|}{\rotatebox{90}{\tabletag{ \textbf{\ref{subsec:when-after}} } After Training}} &
\multicolumn{1}{c}{\rspace \rotatebox{90}{\tabletag{ \textbf{\ref{subsec:where-community}} } Publication Venue}}
\\
\midrule

Abadi, et al., 2016 \cite{abadi2016tensorflow} \rspace & \f & \f & \f & & \f & \f & & & & & & \f & & & \f & & & & \f & \f & \rspace arXiv \\
\rowcolor{rowbackground}
Bau, et al., 2017 \cite{bau2017netdissect} \rspace & \f & & \f & & \f & & & & & \f & & & & & & \f & & \f & & \f & \rspace CVPR \\
Bilal, et al., 2017 \cite{bilal2018convolutional} \rspace & \f & \f & & & \f & & & & & \f & & \f & & & & \f & & \f & \f & & \rspace TVCG \\
\rowcolor{rowbackground}
Bojarski, et al., 2016 \cite{bojarski2016visualbackprop} \rspace & \f & \f & & & \f & & & & \f & & & \f & & & & \f & & \f & \f & \f & \rspace arXiv \\
Bruckner, 2014 \cite{bruckner2014mloscope} \rspace & \f & \f & & & \f & & & \f & \f & & & & \f & & & \f & & \f & \f & & \rspace MS Thesis \\
\rowcolor{rowbackground}
Carter, et al., 2016 \cite{carter2016experiments} \rspace & \f & & & \f & \f & \f & \f & & & \f & \f & \f & & & & \f & \f & & & \f & \rspace Distill \\
Cashman, et al., 2017 \cite{cashman2017rnnbow} \rspace & \f & \f & & & \f & \f & & & \f & \f & & & & & & \f & & & & \f & \rspace VADL \\
\rowcolor{rowbackground}
Chae, et al., 2017 \cite{chae2017visualization} \rspace & \f & \f & & & \f & & & & & \f & & \f & & & \f & \f & & & \f & & \rspace VADL \\
Chung, et al., 2016 \cite{chung2016revacnn} \rspace & \f & \f & & & \f & & & \f & \f & \f & \f & & \f & \f & \f & \f & & & \f & & \rspace FILM \\
\rowcolor{rowbackground}
Goyal, et al., 2016 \cite{goyal2016towards} \rspace & \f & & & & & & \f & & \f & & & & & & & \f & \f & \f & & \f & \rspace arXiv \\
Harley, 2015 \cite{harley2015isvc} \rspace & \f & & & \f & & & \f & \f & \f & \f & & & \f & & & \f & \f & & & \f & \rspace ISVC \\
\rowcolor{rowbackground}
Hohman, et al., 2017 \cite{hohman2017shapeshop} \rspace & \f & & \f & \f & & & \f & & & & & \f & & & & \f & \f & \f & & \f & \rspace CHI \\
Kahng, et al., 2018 \cite{kahng2018activis} \rspace & \f & \f & & & \f & \f & & \f & & \f & \f & \f & \f & \f & & \f & & & & \f & \rspace TVCG \\
\rowcolor{rowbackground}
Karpathy, et al., 2015 \cite{karpathy2015visualizing} \rspace & \f & & & & \f & \f & & & & \f & \f & \f & & \f & & \f & & & & \f & \rspace arXiv \\
Li, et al., 2015 \cite{li2015visualizing} \rspace & \f & & & & \f & \f & & & & \f & \f & \f & & \f & & \f & & & & \f & \rspace arXiv \\
\rowcolor{rowbackground}
Liu, et al., 2017 \cite{liu2017towards} \rspace & \f & \f & & & \f & & & \f & \f & \f & & \f & \f & & & \f & & & & \f & \rspace TVCG \\
Liu, et al., 2018 \cite{liu2018analyzing} \rspace & \f & \f & & & \f & & & \f & \f & \f & & \f & \f & & \f & \f & & & \f & & \rspace TVCG \\
\rowcolor{rowbackground}
Ming, et al., 2017 \cite{ming2017understanding} \rspace & \f & & \f & & \f & & & & & \f & & \f & & & & \f & & & & \f & \rspace VAST \\
Norton \& Qi, 2017 \cite{norton2017adversarial} \rspace & \f & \f & & \f & \f & \f & \f & & & & & & & & & \f & \f & & & \f & \rspace VizSec \\
\rowcolor{rowbackground}
Olah, 2014 \cite{olah2014visualizing} \rspace & \f & & & \f & & & \f & & & & \f & & & \f & & \f & \f & & & \f & \rspace Web \\
Olah, et al., 2018 \cite{olah2018the} \rspace & \f & & & \f & \f & \f & \f & \f & & \f & \f & \f & & & & \f & \f & \f & & \f & \rspace Distill \\
\rowcolor{rowbackground}
Pezzotti, et al., 2017 \cite{pezzotti2017deepeyes} \rspace & \f & \f & & & \f & & & & & \f & \f & \f & & \f & \f & \f & & & \f & & \rspace TVCG \\
Rauber, et al., 2017 \cite{rauber2017visualizing} \rspace & \f & \f & \f & & \f & & & & & \f & \f & \f & & \f & & \f & & & \f & \f & \rspace TVCG \\
\rowcolor{rowbackground}
Robinson, et al., 2017 \cite{robinson2017deeppop} \rspace & \f & & & & \f & \f & & & & \f & \f & \f & & & & \f & & & & \f & \rspace GeoHum. \\
Rong \& Adar, 2016 \cite{rong2016visual} \rspace & \f & \f & & & \f & \f & & & & \f & & \f & & & & \f & & & & \f & \rspace ICML VIS \\
\rowcolor{rowbackground}
Smilkov, et al., 2016 \cite{smilkov2016embedding} \rspace & \f & & & & & \f & & & & \f & \f & \f & & \f & & \f & & & & \f & \rspace NIPS WS. \\
Smilkov, et al., 2017 \cite{smilkov2017direct} \rspace & \f & \f & & \f & & & \f & \f & \f & \f & & & \f & & \f & & \f & & \f & \f & \rspace ICML VIS \\
\rowcolor{rowbackground}
Strobelt, et al., 2018 \cite{strobelt2017lstmvis} \rspace & \f & \f & & & \f & \f & & & & \f & \f & \f & & \f & & \f & & & & \f & \rspace TVCG \\
Tzeng \& Ma, 2005 \cite{tzeng2005opening} \rspace & \f & & & & \f & & & \f & \f & & & \f & \f & & \f & & & & & \f & \rspace VIS \\
\rowcolor{rowbackground}
Wang, et al., 2018 \cite{wang2018ganviz} \rspace & \f & \f & \f & & \f & & & & \f & \f & & \f & & \f & \f & \f & & & \f & & \rspace TVCG \\
Webster, et al., 2017 \cite{webster2017teachable} \rspace & & & & \f & & & \f & & & & & & & & & \f & \f & & \f & \f & \rspace Web \\
\rowcolor{rowbackground}
Wongsuphasawat, et al., 2018 \cite{wongsuphasawat2018visualizing} \rspace & & \f & & & \f & & & \f & & & & \f & \f & & & & & & & & \rspace TVCG \\
Yosinski, et al., 2015 \cite{yosinski2015understanding} \rspace & \f & & & \f & & \f & \f & \f & \f & \f & & & & & & \f & \f & \f & & \f & \rspace ICML DL \\
\rowcolor{rowbackground}
Zahavy, et al., 2016 \cite{zahavy2016graying} \rspace & \f & \f & & & \f & & & & & \f & \f & \f & & \f & & \f & & & & \f & \rspace ICML \\
Zeiler, et al., 2014 \cite{zeiler2014visualizing} \rspace & \f & \f & & & \f & & & & \f & \f & & & & & & & & \f & & \f & \rspace ECCV \\
\rowcolor{rowbackground}
Zeng, et al., 2017 \cite{zeng2017cnncomparator} \rspace & \f & & \f & & \f & & & \f & & \f & & & & & & \f & & & \f & & \rspace VADL \\
Zhong, et al., 2017 \cite{zhong2017evolutionary} \rspace & \f & \f & & & \f & & & & & \f & \f & \f & & \f & \f & \f & & \f & \f & & \rspace ICML VIS \\
\rowcolor{rowbackground}
Zhu, et al., 2016 \cite{zhu2016generative} \rspace & \f & & & & \f & \f & \f & & & & & \f & & & & \f & \f & \f & & \f & \rspace ECCV \\

\end{tabular}
\end{table*}

\begin{table*}[]
\sffamily

\centering

\caption{
Foundational deep learning terminology used in this paper, sorted by importance.
In a term's ``meaning'' (last column), 
defined terms are italicized.
}

\label{table:terminology-foundational}


\setlength{\tabcolsep}{0pt}
\renewcommand\arraystretch{1.5}

\begin{tabular}{P{2.5cm} @{\hskip 0.2cm} P{2.5cm}  @{\hskip 0.2cm} p{12.6cm}}

Technical Term & Synonyms & Meaning \\
\midrule
Neural Network & Artificial neural net, net &
Biologically-inspired models that form the basis of deep learning; approximate functions dependent upon a large and unknown amount of inputs consisting of \textit{layers} of \textit{neurons}\\

Neuron & Computational unit, node &
Building blocks of \textit{neural networks}, entities that can apply \textit{activation functions}\\

Weights & Edges &
The trained and updated parameters in the \textit{neural network} model that connect \textit{neurons} to one another\\

Layer & Hidden layer &
Stacked collection of \textit{neurons} that attempt to extract features from data; a \textit{layer's} input is connected to a previous \textit{layer's} output\\

Computational Graph & Dataflow graph &
Directed graph where nodes represent operations and edges represent data paths; when implementing \textit{neural network} models, often times they are represented as these\\

Activation Functions & Transform function &
Functions embedded into each \textit{layer} of a \textit{neural network} that enable the network represent complex non-linear decisions boundaries \\

Activations & Internal representation &
Given a trained network one can pass in data and recover the \textit{activations} at any \textit{layer} of the network to obtain its current representation inside the network \\

Convolutional Neural Network & CNN, convnet &
Type of \textit{neural network} composed of convolutional \textit{layers} that typically assume image data as input; these \textit{layers} have depth unlike typical \textit{layers} that only have width (number of \textit{neurons} in a \textit{layer}); they make use of filters (feature \& pattern detectors) to extract spatially invariant representations\\

Long Short-Term Memory & LSTM &
Type of \textit{neural network}, often used in text analysis, that addresses the vanishing gradient problem by using memory gates to propagate gradients through the network to learn long-range dependencies\\

Loss Function & Objective function, cost function, error & 
Also seen in general ML contexts, defines what success looks like when learning a representation, i.e., a measure of difference between a \textit{neural network's} prediction and ground truth \\

Embedding & Encoding & 
Representation of input data (e.g., images, text, audio, time series) as vectors of numbers in a high-dimensional space; oftentimes reduced so data points (i.e., their vectors) can be more easily analyzed (e.g., compute similarity)\\

Recurrent Neural Network & RNN &
Type of \textit{neural network} where recurrent connections allow the persistence (or ``memory``) of previous inputs in the network's internal state which are used to influence the network output\\

Generative Adversarial Networks & GAN &
Method to conduct unsupervised learning by pitting a generative network against a discriminative network; the first network mimics the probability distribution of a training dataset in order to fool the discriminative network into judging that the generated data instance belongs to the training set\\

Epoch & Data pass &
A complete pass through a given dataset; by the end of one \textit{epoch}, a \textit{neural network} will have seen every datum within the dataset once\\

\end{tabular}
\end{table*}

\section{Why Visualize Deep Learning}
\label{sec:why}

\subsection{Interpretability \& Explainability}
\label{subsec:why-interpretability}
The most abundant, and to some, the most important reason why people want to visualize deep learning is to understand how deep learning models make decisions and what representations they have learned, so we can place trust in a model~\cite{lipton2016mythos}.
This notion of general model understanding has been called the \textit{interpretability} or \textit{explainability} when referring to machine learning models~\cite{lipton2016mythos, montavon2017methods, miller2017explanation}.
However, neural networks particularly suffer from this problem since oftentimes real world and high-performance models contain a large number of parameters (in the millions) and exhibit extreme internal complexity by using many non-linear transformations at different stages during training.
Many works motivate this problem by using phrases such as ``opening and peering through the black-box,'' ``transparency,'' and ``interpretable neural networks,''~\cite{tzeng2005opening, weller2017challenges, zahavy2016graying}, referring the internal complexity of neural networks.

\subsubsection{Discordant Definitions for Interpretability}
Unfortunately, there is no universally formalized and agreed upon definition for explainability and interpretability in deep learning, which makes classifying and qualifying interpretations and explanations troublesome.
In Lipton's work ``The Mythos of Model Interpretability~\cite{lipton2016mythos},'' he surveys interpretability-related literature, and discovers diverse motivations for why interpretability is important and is occasionally discordant.
Despite this ambiguity, he attempts to refine the notion of interpretability by making a first step towards providing a comprehensive taxonomy of both the desiderata and methods in interpretability research.
One important point that Lipton makes is the difference between interpretability and an explanation; an explanation can show predictions without elucidating the mechanisms by which models work~\cite{lipton2016mythos}.

In another work originally presented as a tutorial at the International Conference on Acoustics, Speech, and Signal Processing by Montavona et al.~\cite{montavon2017methods}, the authors propose exact definitions of both an interpretation and an explanation.
First, an interpretation is ``the mapping of an abstract concept (e.g., a predicted class) into a domain that the human can make sense of.''
They then provide some examples of interpretable domains, such as images (arrays of pixels) and text (sequences of words), and noninterpretable domains, such as abstract vector spaces (word embeddings).
Second, an explanation is ``the collection of features of the interpretable domain, that have contributed for a given example to produce a decision (e.g., classification or
regression).''
For example, an explanation can be a heatmap highlighting which pixels of the input image most strongly support an image classification decision, or in natural language processing, explanations can highlight certain phrases of text.

However, both of the previous works are written by members of the AI community, whereas work by Miller titled ``Explanation in Artificial Intelligence: Insights from the Social Sciences''~\cite{miller2017explanation} postulates that much of the current research uses only AI researchers' intuition of what constitutes a ``good'' explanation.
He suggests that if the focus on explaining decisions or actions to a human observer is the goal, then if these techniques are to succeed, the explanations they generate should have a structure that humans accept.
Much of Miller's work highlights vast and valuable bodies of research in philosophy, psychology, and cognitive science for how people define, generate,
select, evaluate, and present explanations, and he argues that interpretability and explainability research should leverage and build upon this history~\cite{miller2017explanation}.
In another essay, Offert~\cite{offert2017know} argues that to make interpretability more rigorous, we must first identify where it is impaired by intuitive considerations.
That is, we have to ``consider it precisely in terms of what it is not.''
While multiple works bring different perspectives, Lipton makes the keen observation that for the field to progress, the community must critically engage with this problem formulation issue~\cite{lipton2016mythos}.
Further research will help solidify the notions of interpretation and explanation.

\subsubsection{Interpretation as Qualitative Support for Model Evaluation in Various Application Domains}
\label{subsubsec:application}
While research into interpretation itself is relatively new, its impact has already been seen in applied deep learning contexts.
A number of applied data science and AI projects that use deep learning models include a section on interpretation to qualitatively evaluate and support the model's predictions and the work's claims overall.
An example of this is an approach for end-to-end neural machine translation.
In the work by Johnson et al.~\cite{johnson2016google}, the authors present a simple and efficient way to translate between multiple languages using a single model,
taking advantage of multilingual data to improve neural machine translation for all languages involved.
The authors visualize an embedding of text sequences, for example, sentences from multiple languages, to support and hint at a universal interlingua representation.
Another work that visualizes large machine learning embeddings is by Zahavy et al.~\cite{zahavy2016graying}, where the authors analyze deep Q-networks (DQN), a popular reinforcement learning model, to understand and describe the policies learned by DQNs for three different Atari 2600 video games.
An application for social good by Robinson et al.~\cite{robinson2017deeppop} demonstrates how to apply deep neural networks on satellite imagery to perform population prediction and disaggregation, jointly answering the questions ``where do people live?'' and ``how many people live there?''.
In general, they show how their methodology can be an effective tool for extracting information from inherently unstructured, remotely-sensed data to provide effective solutions to social problems.

These are only a few domains where visualization and deep learning interpretation have been successfully used.
Others include building trust in autonomous driving vehicles~\cite{bojarski2016visualbackprop}, explaining decisions made by medical imaging models, such as MRIs on brain scans, to provide medical experts more information for making diagnoses~\cite{zintgraf2017visualizing}, and using visual analytics to explore automatically-learned features from street imagery to gain perspective into identity, function, demographics, and affluence in urban spaces, which is useful for urban design and planning~\cite{li2017hierarchical}.

In this survey we will mention interpretation and explanation often, as they are the most common motivations for deep learning visualization.
Later, we will discuss the different visualization techniques and visual analytics systems that focus on neural network interpretability for embeddings~\cite{smilkov2016embedding}, text~\cite{li2015visualizing, karpathy2015visualizing, carter2016experiments}, quantifying interpretability~\cite{bau2017netdissect}, and many different image-based techniques stemming from the AI communities~\cite{erhan2009visualizing, zeiler2014visualizing, selvaraju2016grad, simonyan2013deep, nguyen2016synthesizing}.

\subsection{Debugging \& Improving Models}
\label{subsec:why-debugging}
Building machine learning models is an iterative design process~\cite{patel2008investigating, kulesza2015principles, nushi2017human}, and developing deep neural networks is no different.
While mathematical foundations have been laid, deep learning still has many open research questions.
For example, finding the exact combinations of model depth, layer width, and finely tuned hyperparameters is nontrivial.
In response to this, many visual analytics systems have been proposed to help model developers build and debug their models, with the hope of expediting the iterative experimentation process to ultimately improve performance~\cite{wongsuphasawat2018visualizing, strobelt2017lstmvis, pezzotti2017deepeyes}.
Oftentimes this requires monitoring models during the training phase~\cite{zhong2017evolutionary, liu2018analyzing}, identifying misclassified instances and testing a handful of well-known data instances to observe performance~\cite{kahng2018activis, bilal2018convolutional, rong2016visual}, 
and allowing a system to suggest potential directions for the model developer to explore~\cite{chae2017visualization}.
This reason for why we wish to visualize deep learning ultimately provides better tools to speed up model development for engineers and researchers so that they can quickly identify and fix problems within a model to improve  overall performance.

\subsection{Comparing \& Selecting Models}
\label{subsec:why-comparing}
While certainly related to model debugging and improvement, model comparison and selection are slightly different tasks in which visualization can be useful~\cite{alexander2016task, mcmahan2013ad, kahng2016visual}.
Oftentimes model comparison describes the notion of choosing a single model among an ensemble of well-performing models. That is, no debugging needs to be done; all models have ``learned'' or have been trained semi-successfully.
Therefore, the act of selecting a single, best-performing model requires inspecting model metrics and visualizing parts of the model to pick the one that has the highest accuracy, the lowest loss, or is the most generalizable, while avoiding pitfalls such as memorizing training data or overfitting.

Some systems take a high-level approach and compare user-defined model metrics, like accuracy and loss, and aggregate them on interactive charts for  performance comparison~\cite{abadi2016tensorflow}.
Other frameworks compare neural networks trained on different random initializations (an important step in model design) to discover how they would affect performance, while also quantifying performance and interpretation~\cite{bau2017netdissect}.
Some approaches compare models on image generation techniques, such as performing image reconstruction from the internal representations of each layer of different networks to compare different network architectures~\cite{yu2014visualizing}.
Similar to comparing model architectures, some systems solely rely on data visualization representations and encodings to compare models~\cite{ming2017understanding}, while others compare different snapshots of a single model as it trains over time, i.e., comparing a model after $n_1$ epochs and the same model after $n_2$ epochs of training time~\cite{zeng2017cnncomparator}.

\subsection{Teaching Deep Learning Concepts}
\label{subsec:why-teaching}
Apart from AI experts, another important reason why we may wish to visualize deep learning is to educate non-expert users about AI.
The exact definition of non-experts varies by source and is discussed further in Section \ref{subsec:who-nonexpert}.
An example that targets the general public is Teachable Machines~\cite{webster2017teachable}, a web-based AI experiment that explores and teaches the foundations of an image classifier.
Users train a three-way image classifier by using their computer's webcam to generate the training data.
After providing three different examples of physical objects around the user (e.g., holding up a pencil, a coffee mug, and a phone), the system then performs real-time inference on whichever object is in view of the webcam, and shows a bar chart with the corresponding classification scores.
Since inference is computed in real-time, the bar charts wiggles and jumps back and forth as the user removes an object, say the pencil, from the view and instead holds up the coffee mug.
The visualization used is a simple bar chart, which provides an approachable introduction into image classification, a modern-day computer vision and AI problem.

Another example for teaching deep learning concepts, the Deep Visualization Toolbox~\cite{yosinski2015understanding} discussed later in this survey, also uses a webcam for instant feedback when interacting with a neural network.
Taking instantaneous feedback a step further, some works have used direct manipulation to engage non-experts in the learning process.
TensorFlow Playground~\cite{smilkov2017direct}, a robust, web-based visual analytics tool for exploring simple neural networks, uses direct manipulation to reinforce deep learning concepts, and importantly, evokes the user's intuition about how neural networks work.
Other non-traditional mediums have been used to teach deep learning concepts and build an intuition for how neural networks behave too.
Longform, interactive scrollytelling works focusing on particular AI topics that use interactive visualizations as supporting evidence are gaining popularity.
Examples include ``How to Use t-SNE Effectively,'' where users can play with hundreds of small datasets and vary single parameters to observe their effect on an embedding
~\cite{wattenberg2016how}, and a similar interactive article titled ``Visualizing MNIST'' that visualizes different types of embeddings produced by different algorithms~\cite{olah2014visualizing}.

\section{Who Uses Deep Learning Visualization}
\label{sec:who}

This section describes the groups of people who may stand to benefit from deep learning visualization and visual analytics.
We loosely organize them into three \textit{non-mutually exclusive} groups by their level of deep learning knowledge (most to least): \textit{model developers}, \textit{model users}, and \textit{non-experts}.
Note that many of the works discussed can benefit multiple groups, e.g., a model developer may use a tool aimed at non-experts to reinforce their own intuition for how neural networks learn.

\subsection{Model Developers \& Builders}
\label{subsec:who-developer}
The first group of people who use deep learning visualization are individuals whose job is primarily focused on developing, experimenting with, and deploying deep neural networks.
These model developers and builders, whether they are researchers or engineers, have a strong understanding of deep learning techniques and a well-developed intuition surrounding model building.
Their knowledge expedites key decisions in deep learning workflows, such as identifying the which types of models perform best on which types of data.
These individuals wield mastery over models, e.g., knowing how to vary hyperparameters in the right fashion to achieve better performance.
These individuals are typically seasoned in building large-scale models and training them on high-performance machines to solve real-world problems~\cite{liu2017towardsva}.
Therefore, tooling and research for these users is much more technically focused, e.g., exposing many hyperparameters for detailed model control.

Of the existing deep learning visual analytics tools published, a handful tackle the problem of developing tools for model developers, but few have seen widespread adoption.
Arguably the most well-known system is TensorBoard~\cite{abadi2016tensorflow}: Google's included open-source visualization platform for its dataflow graph library TensorFlow.
TensorBoard includes a number of built-in components to help model developers understand, debug, and optimize TensorFlow programs.
It includes real-time plotting of quantitative model metrics during training, instance-level predictions, and a visualization of the computational graph.
The computational graph component was published separately by Wongsuphasawat et al.~\cite{wongsuphasawat2018visualizing} and works by applying a series of graph transformations that enable standard layout techniques to produce interactive diagrams of TensorFlow models.

Other tools, such as DeepEyes~\cite{pezzotti2017deepeyes}, assist in a number of model building tasks, e.g., 
identifying stable layers during the training process,
identifying unnecessary layers and degenerated filters that do not contribute to a model's decisions, 
pruning such entities, 
and identifying patterns undetected by the network, indicating that more filters or layers may be needed.
Another tool, \textit{Blocks}~\cite{bilal2018convolutional}, allows a model builder to accelerate model convergence and alleviate overfitting, through visualizing class-level confusion patterns.
Other research has developed new metrics beyond measures like loss and accuracy, to help developers inspect and evaluate networks while training them~\cite{zhong2017evolutionary}.

Some tools also address the inherent iterative nature of training neural networks.
For example, ML-o-scope~\cite{bruckner2014mloscope} utilizes a time-lapse engine to inspect a model's training dynamics to better tune hyperparameters, while work by Chae et al.~\cite{chae2017visualization} visualizes classification results during training and suggests potential directions to improve performance in the model building pipeline.
Lastly, visual analytics tools are beginning to be built for expert users who wish to use  models that are more  challenging to work with.
For example, DGMTracker~\cite{liu2018analyzing} is a visual analytics tool built to help users understand and diagnose the training process of deep generative models:  powerful networks that perform unsupervised and semi-supervised learning where the primary focus is to discover the hidden structure of data without resorting to external labels.

\begin{figure}[tb]
 \centering
 \includegraphics[width=\columnwidth]{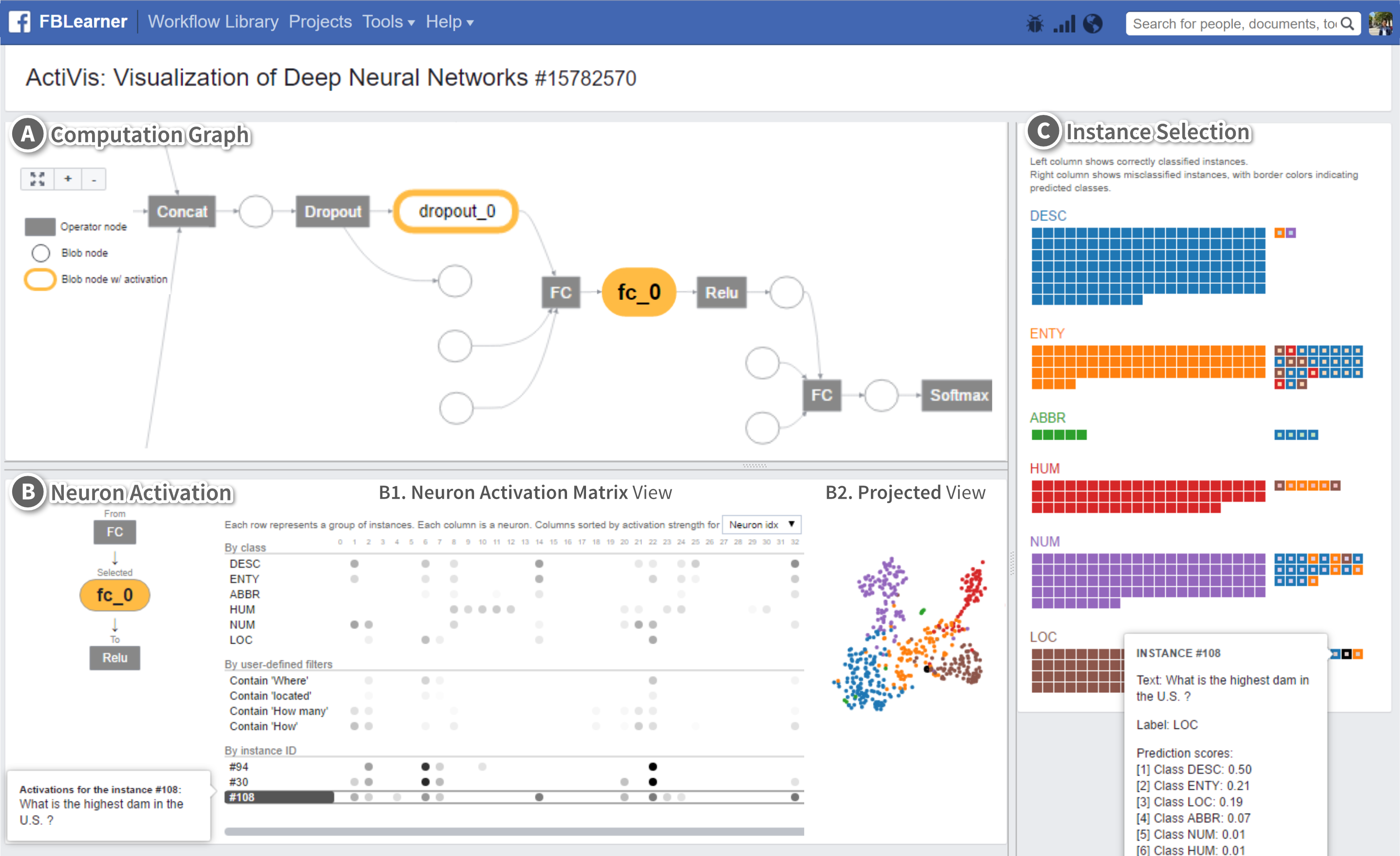}
 \caption{
  ActiVis~\cite{kahng2018activis}: a visual analytics system for interpreting neural network results using a novel visualization that unifies instance- and subset-level inspections of neuron activations deployed at Facebook.
 }
 \label{fig:activis}
\end{figure}

\subsection{Model Users}
\label{subsec:who-user}
The second group of people who may benefit from deep learning visualization are model users.
These are users who may have some technical background but are neural network novices.
Common tasks include using well-known neural network architectures for developing domain specific applications, training smaller-scale models, and downloading pre-trained model weights online to use as a starting point.
This group of users also include machine learning artists who use models to enable and showcase new forms of artistic expression.

An example visual analytics system for these model users is ActiVis~\cite{kahng2018activis}: a visual analytics system for interpreting the results of neural networks by using a novel visual representation that unifies instance- and subset-level inspections of neuron activations.
Model users can flexibly specify subsets using input features, labels, or any intermediate
outcomes in a machine learning pipeline.
ActiVis was built for engineers and data scientists at Facebook to explore and interpret deep learning models results and is deployed on Facebook's internal system.
LSTMVis~\cite{strobelt2017lstmvis} is a visual analysis tool for recurrent neural networks with a focus on understanding hidden state dynamics in sequence modeling.
The tool allows model users to perform hypothesis testing by selecting an input range to focus on local state changes, then to match these states changes to similar patterns in a large dataset, and finally align the results with structural annotations.
The LSTMVis work describes three types of users: 
architects, those who wish to develop new deep learning methodologies; 
trainers, those who wish to apply LSTMs to a task in which they are domain experts in; and 
end users, those who use pretrained models for various tasks.
Lastly, Embedding Projector~\cite{smilkov2016embedding}, while not specifically deep learning exclusive, is a visual analytics tool to support interactive visualization and interpretation of large-scale embeddings, which are common outputs from neural network models.
The work presents three important tasks that model users often perform while using embeddings; these include exploring local neighborhoods, viewing the global geometry to find clusters, and finding meaningful directions within an embedding.

\subsection{Non-experts}
\label{subsec:who-nonexpert}
The third group of people whom visualization could aid are non-experts in deep learning.
These are individuals who typically have no prior knowledge about deep learning, and may or may not have a technical background.
Much of the research targeted at this group is for educational purposes, trying to explain what a neural network is and how it works at a high-level, sometimes without revealing deep learning is present.
These group also includes people who simply use AI-powered devices and consumer applications.

Apart from Teachable Machines~\cite{webster2017teachable} and the Deep Visualization Toolbox~\cite{yosinski2015understanding} mentioned in Section~\ref{subsec:why-teaching}, TensorFlow Playground~\cite{smilkov2017direct}, a web-based interactive visualization of a simple dense network, has become a go-to tool for gaining intuition about how neural networks learn.
TensorFlow Playground uses direct manipulation experimentation rather than coding, enabling users to quickly build an intuition about neural networks.
The system has been used to teach students about foundational neural network properties by using ``living lessons,'' and also makes it straightforward to create a dynamic, interactive educational experience.
Another web-browser based system, ShapeShop~\cite{hohman2017shapeshop}, allows users to explore and understand the relationship between input data and a network's learned representations.
ShapeShop uses a feature visualization technique called class activation maximization to visualize specific classes of an image classifier.
The system allows users to interactively select classes from a collection of simple shapes, select a few hyperparameters, train a model, and view the generated visualizations all in real-time.

Tools built for non-experts, particularly with an educational focus, are becoming more popular on the web.
A number of web-based JavaScript frameworks for training neural networks and inference have been developed; however, ConvNetJS (\url{http://cs.stanford.edu/people/karpathy/convnetjs/}) and TensorFlow.js (\url{https://js.tensorflow.org/}) are the most used and have enabled developers to create highly interactive explorable explanations for deep learning models.

\section{What to Visualize in Deep Learning}
\label{sec:what}

This section discusses the technical components of neural networks that could be visualized.
This section is strongly related to the next section, Section \ref{sec:how} ``How,'' which describes how the components of these networks are visualized in existing work.
By first describing \textit{what} may be visualized (this section), we can more easily ground our discussion on \textit{how} to visualize them (next section).

\subsection{Computational Graph \& Network Architecture}
\label{subsec:what-architecture}
The first thing that can be visualized in a deep learning model is the model architecture.
This includes the \textit{computational graph} that defines how a neural network model would train, test, save data to disk, and checkpoint after epoch iterations~\cite{abadi2016tensorflow}.
Also called the dataflow graph~\cite{abadi2016tensorflow}, this defines how data flows from operation to operation to successfully train and use a model.
This is different than the neural network's edges and weights, discussed next, which are the parameters to be tweaked during training.
The dataflow graph can be visualized to potentially inform model developers of the types of computations occurring within their model, as discussed in Section \ref{subsec:how-architecture}.

\subsection{Learned Model Parameters}
\label{subsec:what-parameters}
Other components that can be visualized are the learned parameters in the network during and after training.

\subsubsection{Neural Network Edge Weights}
Neural network models are built of many, and sometimes diverse, constructions of layers of computational units~\cite{Goodfellow-et-al-2016}.
These layers send information throughout the network by using edges that connect layers to one another, oftentimes in a linear manner, yet some more recent architectures have shown that skipping certain layers and combining information in unique ways can lead to better performance. 
Regardless, each node has an outgoing edge with an accompanying \textit{weight} that sends signal from one neuron in a layer to potentially thousands of neurons in an adjacent layer~\cite{smilkov2017direct}.
These are the parameters that are tweaked during the backpropagation phase of training a deep model, and could be worthwhile to visualize for understanding what the model has learned, as seen in Section \ref{subsec:how-architecture}.

\subsubsection{Convolutional Filters}
Convolutional neural networks are built using a particular type of layer, aptly called the \textit{convolutional layer}.
These convolutional layers apply filters over the input data, oftentimes images represented as a two-dimensional matrix of values, to generate smaller representations of the data to pass to later layers in the network.
These filters, like the previously mentioned traditional weights, are then updated throughout the training process, i.e., learned by the network, to support a given task.
Therefore, visualizing the learned filters could be useful as an alternate explanation for what a model has learned \cite{zeiler2014visualizing, yosinski2015understanding}, as seen in Section \ref{subsec:how-algorithms}.

\subsection{Individual Computational Units}
\label{subsec:what-individual}
Albeit reductionist, neural networks can be thought as a collection of layers of neurons connected by edge weights.
Above, we discussed that the edges can be visualized, but the neurons too can be a source of data to investigate.

\subsubsection{Activations}
When given a trained model, one can perform inference on the model using a new data instance to obtain the neural network's output, e.g., a classification or a specific predicted value.
Throughout the network, the neurons compute \textit{activations} using activation functions (e.g., weighted sum) to combine the signal from the previous layer into a new node~\cite{Goodfellow-et-al-2016, yosinski2015understanding}.
This mapping is one of the characteristics that allows a neural network to learn.
During inference, we can recover the activations produced at each layer.
We can use activations in multiple ways, e.g., as a collection of individual neurons, spatial positions, or channels~\cite{olah2018the}.
Although these feature representations are typically high-dimensional vectors of the input data at a certain stage within the network~\cite{olah2018the}, it could be valuable in helping people visualize how input data is transformed into higher-level features, as seen in Section \ref{subsec:how-scatter}.
Feature representations may also shed light upon how the network and its components respond to particular data instances \cite{yosinski2015understanding}, commonly called instance-level observation; we will discuss this in detail in Section \ref{subsec:how-instance} and  \ref{subsec:how-ie}.

\subsubsection{Gradients for Error Measurement}
To train a neural network, we commonly use a process known as backpropagation~\cite{Goodfellow-et-al-2016}.
\textit{Backpropagation}, or sometimes called the backpropagation of errors, is a method to calculate the gradient of a specified loss function.
When used in combination with an optimization algorithm, e.g., gradient descent, we can compute the error at the output layer of a neural network and redistribute the error by updating the model weights using the computed gradient.
These gradients flow over the same edges defined in the network that contain the weights, but flow in the opposite direction., e.g., from the output layer to the input layer.
Therefore, it could be useful to visualize the gradients of a network to see how much error is produced at certain outputs and where it is distributed~\cite{chung2016revacnn, cashman2017rnnbow}, as mentioned in Section \ref{subsec:how-algorithms}.

\subsection{Neurons in High-dimensional Space}
\label{subsec:what-highdimensional}
Continuing the discussion of visualizing activations of a data instance, we can think of the feature vectors recovered as vectors in a high-dimensional space.
Each neuron in a layer then becomes a ``dimension.''
This shift in perspective is powerful, since we can now take advantage of high-dimensional visualization techniques to visualize extracted activations~\cite{rauber2017visualizing, maaten2008visualizing}.
Sometimes, people use neural networks simply as feature vector generators, and defer the actual task to other computational techniques, e.g., traditional machine learning models~\cite{robinson2017deeppop, simonyan2013deep}.
In this perspective, we now can think of deep neural networks as feature generators, whose output embeddings could be worth exploring.
A common technique is to use dimensionality reduction to take the space spanned by the activations and embed it into 2D or 3D for visualization purposes~\cite{maaten2008visualizing, smilkov2016embedding, rauber2017visualizing}, as discussed in Section \ref{subsec:how-scatter}.

\subsection{Aggregated Information}
\label{subsec:what-aggregated}

\subsubsection{Groups of Instances}
As mentioned earlier, instance-level activations allow one to recover the mapping from data input to a feature vector output.
While this can be done for a single data instance, it can also be done on collections of instances.
While at first this does not seem like a major differentiation from before, instance groups provide some unique advantages~\cite{kahng2018activis, ming2017understanding}.
For example, since instance groups by definition are composed of many instances, one can compute all the activations simultaneously.
Using visualization, we can now compare these individual activations to see how similar or different they are from one another.
Taking this further, with instance groups, we can now take multiple groups, potentially from differing classes, and compare how the distribution of activations from one group compares or differs from another.
This aggregation of known instances into higher-level groups could be useful for uncovering the learned decision boundary in classification tasks, as seen in Section \ref{subsec:how-scatter} and Section \ref{subsec:how-instance}.

\subsubsection{Model Metrics}
\label{subsec:what-metrics}
While instance- and group-level activations could be useful for investigating how neural networks respond to particular results a-priori, they suffer from scalability issues, since deep learning models typically wrangle large datasets.
An alternative object to visualize are model metrics, including loss, accuracy, and other measures of error~\cite{abadi2016tensorflow}.
These summary statistics are typically computed every epoch and represented as a time series over the course of a model's training phase.
Representing the state of a model through a single number, or handful of numbers, abstracts away much of the subtle and interesting features of deep neural networks; however, these metrics are key indicators for communicating how the network is progressing during the training phase~\cite{pezzotti2017deepeyes}.
For example, is the network ``learning'' anything at all or is it learning ``too much'' and is simply memorizing data causing it to overfit?
Not only do these metrics describe notions of a single model's performance over time, but in the case of model comparison, these metrics become more important, as they can provide a quick and easy way to compare multiple models at once.
For this reason, visualizing model metrics can be an important and powerful tool to consider for visual analytics, as discussed in Section \ref{subsec:how-line}.

\section{How to Visualize Deep Learning}
\label{sec:how}

In the previous section, we described what technical components of neural networks could be visualized.
In this section, we summarize how the components are visualized and interacted with in existing literature.
For most neural network components, they are often visualized using a few common approaches.
For example, network architectures are often represented as node-link diagrams;
embeddings of many activations are typically represented as scatter plots; 
and model metrics over epoch time are almost always represented as line charts.
In this section, we will also discuss other representations, going beyond the typical approaches.

\subsection{Node-link Diagrams for Network Architectures}
\label{subsec:how-architecture}
Given a neural network's dataflow graph or model architecture, the most common way to visualize where data flows and the magnitude of edge weights is a node-link diagram.
Neurons are shown as nodes, and edge weights as links.
For computational and dataflow graphs, Kahng et al.~\cite{kahng2018activis} describe two methods for creating node-link diagrams.
The first represents only operations as nodes, while the second represents both operations and data as nodes.
The first way is becoming the standard due to the popularity of TensorBoard~\cite{abadi2016tensorflow} and the inclusion of its interactive dataflow graph visualization~\cite{wongsuphasawat2018visualizing}.
However, displaying large numbers of links from complex models can generate ``hairball'' visualizations where many edge crossings impede pattern discovery. 
To address this problem, Wongsuphasawat et al.~\cite{wongsuphasawat2018visualizing} extracts high-degree nodes (responsible for many of the edge crossings), visualizes them separately from the main graph,
and allow users to define super-groups within the code.
Another approach to reduce clutter is to place more information on each node;  DGMTracker~\cite{liu2018analyzing} provides a quick snapshot of the dataflow in and out of a node by visualizing its activations within each node.

Regarding neural network architecture, many visual analytics systems use node-link diagrams (neurons as nodes,  weights as links)~\cite{tzeng2005opening,smilkov2017direct,harley2015isvc,liu2017towards,chung2016revacnn}.
The weight magnitude and sign can then be encoded using color or link thickness.
This technique was one of the the first to be proposed~\cite{tzeng2005opening}, and the trend has continued on in literature.
Building on this technique, Harley~\cite{harley2015isvc} visualizes the convolution windows on each layer and how the activations propagate through the network to make the final classification.
Similar to the dataflow graph examples above, some works include richer information inside each node besides an activation value, such as showing a list of images that highly activate that neuron or the activations at a neuron as a matrix~\cite{liu2017towards}.
As mentioned in the dataflow graph visualizations, node-link diagrams for network architecture work well for smaller networks~\cite{smilkov2017direct}, but they also suffer from scalabilty issues.
CNNVis~\cite{liu2017towards}, a visual analytics system that visualizes convolutional neural networks, proposes to use a bi-clustering-based edge bundling technique to reduce visual clutter caused by too many links.

\begin{figure}[tb]
 \centering
 \includegraphics[width=\columnwidth]{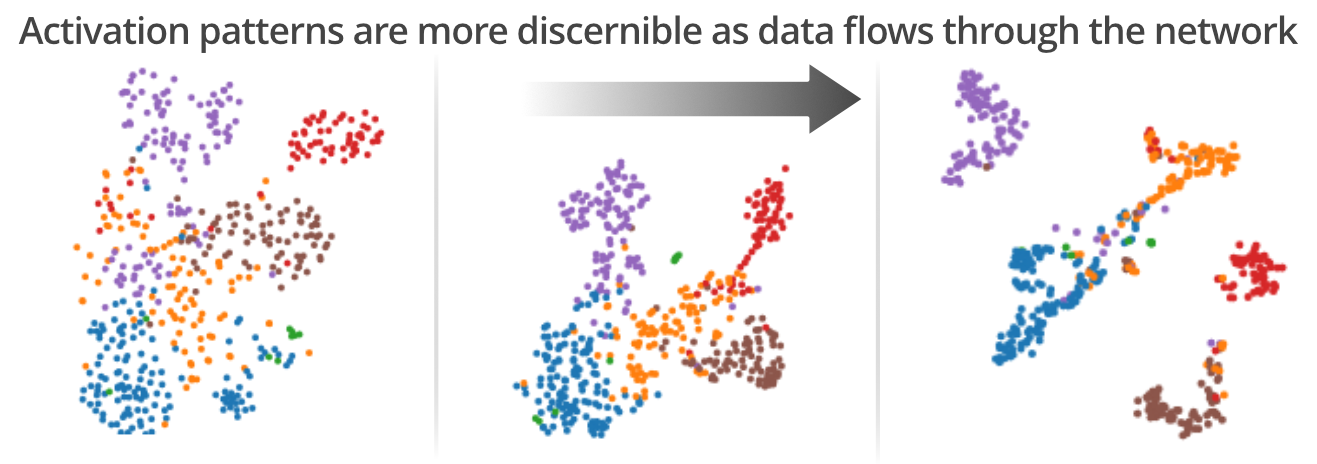}
 \caption{
 Each point is a data instance's high-dimensional activations at a particular layer inside of a neural network, dimensionally reduced, and plotted in 2D.
 Notice as the data flows through the network the activation patterns become more discernible (left to right)~\cite{kahng2018activis}.
 }
 \label{fig:playground}
\end{figure}

\subsection{Dimensionality Reduction \& Scatter Plots}
\label{subsec:how-scatter}
In Section \ref{sec:what}, ``What,'' we discussed different types of high-dimensional embeddings: text can be represented as vectors in word embeddings for natural language processing and images can be represented as feature vectors inside of a neural network.
Both of these types of embeddings are mathematically represented as large tensors, or sometimes as 2D matrices, where each row may correspond to an instance and each column a feature.

The most common technique to visualize these embeddings is performing dimensionality reduction to reduce the number of columns (e.g., features) to two or three.
Projecting onto two dimensions would mean computing $(x,y)$ coordinates for every data instance; for three dimensions, we compute an additional $z$ component, resulting in $(x,y,z)$.
In the 2D case, we can plot all data instances as points in a scatter plot where the axes may or may not have interpretable meaning, depending on the reduction technique used, e.g., principal component analysis (PCA) or t-distributed stochastic neighbor embedding (t-SNE)~\cite{maaten2008visualizing}.
In the 3D case, we can still plot each data instance as a point in 3D space and use interactions to pan, rotate, and navigate this space~\cite{smilkov2016embedding}.
These types of embeddings are often included in visual analytics systems as one of the main views~\cite{pezzotti2017deepeyes, chung2016revacnn}, and are also used in application papers as static figures~\cite{johnson2016google, zahavy2016graying}.
However, viewing a 3D space on a 2D medium (e.g., computer screen) may not be ideal for tasks like comparing exact distances.

Since each reduced point corresponds to an original data instance, another common approach is to retrieve the original image and place it at the reduced coordinate location.
Although the image size must be greatly reduced to prevent excessive overlap, viewing all the images at once can provide insight into what a deep learning model has learned, as seen in the example in \cite{yu2014visualizing} where the authors visualize ImageNet test data, or in \cite{nguyen2016multifaceted} where the authors create many synthetic images from a single class and compare the variance across many random initial starting seeds for the generation algorithm.
We have discussed the typical case where each dot in the scatter plot is a data instance, but some work has also visualized neurons in a layer as separate data instances~\cite{zhong2017evolutionary}.
Another work studies closely how data instances are transformed as their information is passed through the deep network, which in effect visualizes how the neural network separates various classes along approximated decision boundaries~\cite{rauber2017visualizing}.
It is also possible to use time-dependent data and visualize how an embedding changes over time, or in the case of deep learning, over epochs~\cite{rauber2016visualizing}.
This can be useful for evaluating the quality of the embedding during the training phase.

However, these scatter plots raise problems too.
The quality of the embeddings greatly depends on the algorithm used to perform the reduction.
Some works have studied how PCA and t-SNE differ, mathematical and visually, and suggest new reduction techniques to capture the semantic and syntactic qualities  within word embeddings~\cite{liu2018visual}.
It has also been shown that popular reduction techniques like t-SNE are sensitive to changes in the hyperparameter space.
Wattenberg meticulously explores the hyperparameter space for t-SNE, and offers lessons learned and practical advice for those who wish to use dimensionality reduction methods~\cite{wattenberg2016how}.
While these techniques are commonplace, there are still iterative improvements that can be done using clever interaction design, such as finding instances similar to a target instance, i.e., those ``near'' the target in the projected space, helping people build intuition for how data is spatially arranged~\cite{smilkov2016embedding}.

\subsection{Line Charts for Temporal Metrics}
\label{subsec:how-line}
Model developers track the progression of their deep learning models by monitoring and observing a number of different metrics computed after each epoch, including the loss, accuracy, and different measure of errors.
This can be useful for diagnosing the long training process of deep learning models.,
The most common visualization technique for visualizing this data is by considering the metrics as time series and plotting them in line charts~\cite{abadi2016tensorflow}. 
This approach is widely used in deep learning visual analytics tools~\cite{pezzotti2017deepeyes, chung2016revacnn}.
After each epoch, a new entry in the time series is computed, therefore some tools, like TensorBoard, run alongside models as they train and update with the latest status~\cite{abadi2016tensorflow}.
TensorBoard focuses much of its screen real-estate to these types of charts and supports interactions for plotting multiple metrics in small multiples, plotting multiple models on top of one another, filtering different models, providing tooltips for the exact metric values, and resizing charts for closer inspection.
This technique appears in many visual analytics systems and has become a staple for model training, comparison, and selection.

\subsection{Instance-based Analysis \& Exploration}
\label{subsec:how-instance}
Another technique to help interpret and debug deep learning models is testing specific data instances to understand how they progress throughout a model.
Many experts have built up their own collection of data instances over time, having developed deep knowledge about their expected behaviors in models while also knowing their ground truth labels~\cite{kahng2018activis, ren2017squares}.
For example, an instance consisting of a single image or a single text phrase is much easier to understand than an entire image dataset or word embedding consisting of thousands of numerical features extracted from an end user's data.
This is called instance-level observation, where intensive analysis and scrutiny is placed on a single data instance's transformation process throughout the network, and ultimately its final output.

\subsubsection{Identifying \& Analyzing Misclassified Instances}
One application of instance-level analysis is using instances as unit tests for deep learning models.
In the best case scenario, all the familiar instances are classified or predicted correctly; however, it is important to understand \textit{when} a specific instance can fail and \textit{how} it fails.
For example, in the task of predicting population from satellite imagery, the authors showcase three maps of areas with high errors by using a translucent heatmap overlaid on the satellite imagery~\cite{robinson2017deeppop}.
Inspecting these instances reveals three geographic areas that contain high amounts of man-made features and signs of activity but have no registered people living in them: an army base, a national lab, and Walt Disney World. The visualization helps demonstrate that the proposed model is indeed learning high-level features about the input data.
Another technique, HOGgles~\cite{vondrick2013hoggles}, uses an algorithm to visualize feature spaces by using object detectors while inverting visual features back to natural images.
The authors find that when visualizing the features of misclassified images, although  the classification is wrong in the image space, they look deceptively similar to the true positives in the feature space.
Therefore, by visualizing feature spaces of misclassified instances, we can gain a more intuitive understanding of recognition systems.

For textual data, a popular technique for analyzing particular data instances is to use color as the primary encoding.
For example, the background of particular characters in a phrase of words in a sentence would be colored using a divergent color scheme according to some criteria, often their activation magnitudes~\cite{karpathy2015visualizing, carter2016experiments, goyal2016towards}.
This helps identify particular data instances that may warrant deeper inspection (e.g., those misclassified)~\cite{ren2017squares}.

When pre-defined data instances are not unavailable (e.g., when analyzing a new dataset), 
how can we guide users towards important and interesting instances?
To address this problem, a visual analytics system called \textit{Blocks}~\cite{bilal2018convolutional} uses confusion matrices, a technique for summarizing the performance of a classification algorithm, and matrix-level sorting interactions to reveal that class error often occurs in hierarchies.
\textit{Blocks} incorporates these techniques with a sample viewer in the user interface to show selected samples potentially worth exploring.

\subsubsection{Analyzing Groups of Instances}
Instead of using individual data instances for testing and debugging a model, it is also common for experts to perform similar similar tasks using groups of instances~\cite{ren2017squares}.
While some detail may be lost when performing group-level analysis it allows experts to further test the model by evaluating its average and aggregate performance across different groups.

Much of the work using this technique is done on text data using LSTM models~\cite{strobelt2017lstmvis}.
Some approaches compute the saliency for groups of words across the model and visualize the values as a matrix~\cite{li2015visualizing}, while others use matrix visualizations to show the activations of word groups when represented as feature vectors in word embeddings~\cite{rong2016visual, rong2014word2vec}.
One system, ActiVis~\cite{kahng2018activis}, places instance group analysis at the focus of its interactive interface, allowing users to compare preset and user-defined groups of activations.
Similar to the matrix visualization that summarizes activations for each class in CNNVis~\cite{liu2017towards}, ActiVis also uses a scrolling matrix visualization to unify both instance-level and group-level analysis into a single view where users can compare the activations of the user-defined instances.

However, sometimes it can be challenging to define groups for images or text.
For textual data, people often use words to group documents and provide aggregated data.
ConceptVector~\cite{park2018conceptvector} addresses the instance group generation problem by providing an interactive interface to create interesting groups of concepts for model testing.
Furthermore, this system also suggests additional words to include in the user-defined groups, helping guide the user to create semantically sound concepts.

\subsection{Interactive Experimentation}
\label{subsec:how-ie}
Interactive experimentation, another interesting area that integrates deep learning visualization, makes heavy use of interactions for users to experiment with models~\cite{weld2018intelligible}.
By using direct manipulation for testing models, a user can pose ``what if?'' questions and observe how the input data affects the results.
Called \textit{explorable explanations}~\cite{olah2017research}, this type of visual experimentation is popular for 
making sense of complex concepts and systems.

\subsubsection{Models Responding to User-provided Input Data}
To engage the user with the desired concepts to be taught, many systems require the user to provide some kind of input data into the system to obtain results.
Some visual analytics systems use a webcam to capture live videos, and visualize how the internals of neural network models respond to these dynamic inputs~\cite{yosinski2015understanding}.
Another example is a 3D visualization of a CNN trained on the classic MNIST dataset
\footnote{MNIST is a small, popular dataset consisting of thousands of 28$\times$28px images of handwritten digits (0 to 9). MNIST is commonly used as a benchmark for image classification models.}
that shows the convolution windows and activations on images that the user draws by hand~\cite{harley2015isvc}.
For example, drawing a ``5'' in the designated area passes that example throughout the network and populates the visualization with the corresponding activations using a node-link diagram.
A final example using image data is ShapeShop~\cite{hohman2017shapeshop}, a system that allows a user to select data from a bank of simple shapes to be classified.
The system then trains a neural network and using the class activation maximization technique to generate visualizations of the learned features of the model.
This can be done in real-time, therefore a user can quickly train multiple models with different shapes to observe the effect of adding more diverse data to improve the internal model representation.

\begin{figure}[tb]
 \centering
 \includegraphics[width=\columnwidth]{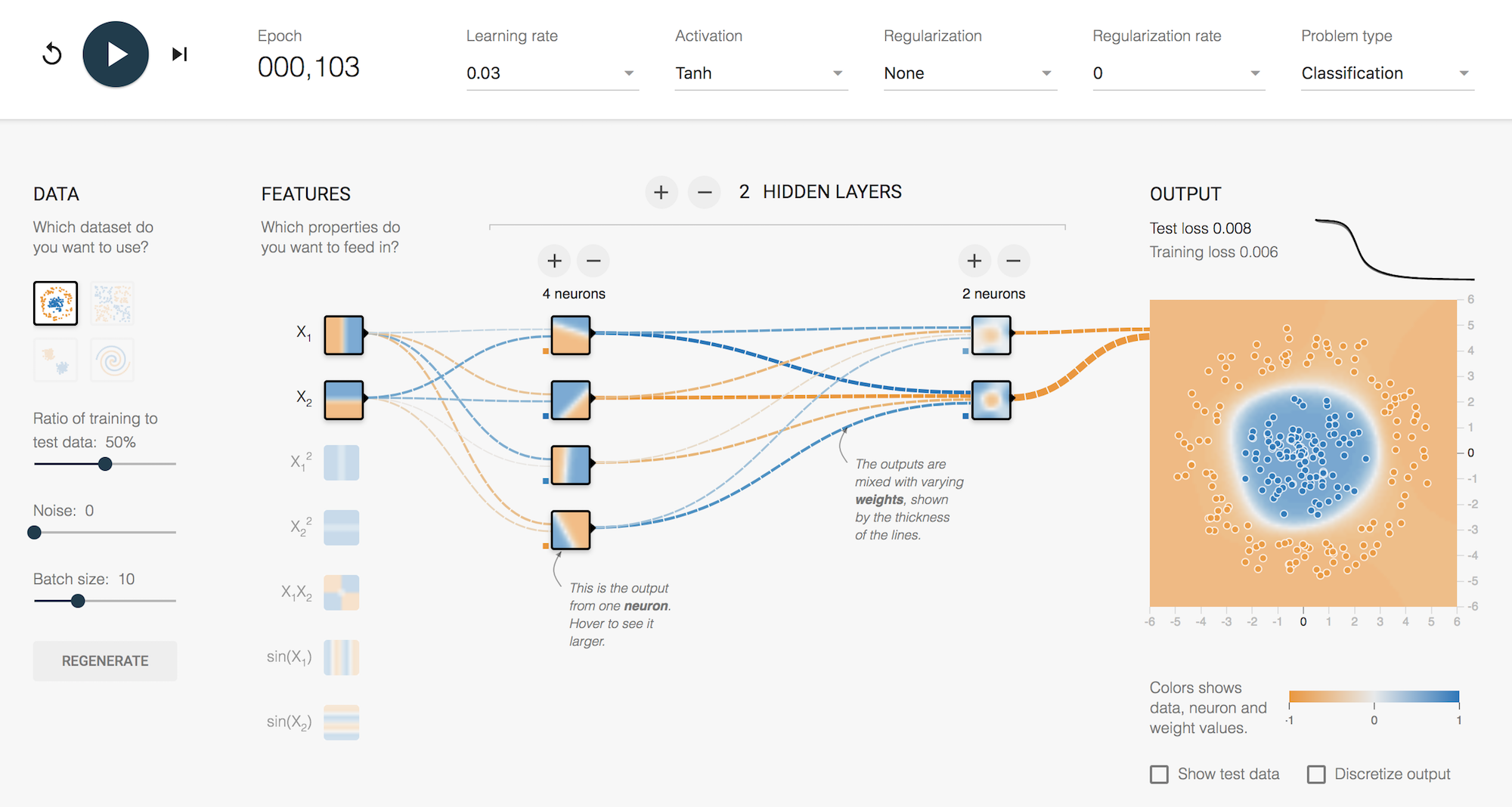}
 \caption{
 TensorFlow Playground~\cite{smilkov2017direct}: a web-based visual analytics tool for exploring simple neural networks that uses direct manipulation rather than programming to teach deep learning concepts and develop an intuition about how neural networks behave.
 }
 \label{fig:playground}
\end{figure}

An example using textual data is the online, interactive Distill article for handwriting prediction~\cite{carter2016experiments}, which allows a user to write words on screen, and in real-time, the system draws multiple to-be-drawn curves predicting what the user's next stroke would be, while also visualizing the model's activations.
Another system uses GANs to interactively generate images based off of user's sketches~\cite{zhu2016generative}.
By sketching a few colored lines, the system presents the user with multiple synthetic images using the sketch as a guideline for what to generate.
A final example is the Adversarial Playground~\cite{norton2017adversarial}, a visual analytics system that enables users to compare adversarially-perturbed images, to help users understand why an adversarial example can fool a CNN image classifier.
The user can select from one of the MNIST digits and adjust the strength of adversarial attack.
The system then compares the classifications scores in a bar chart to observe how simple perturbations can greatly impact classification accuracy.

\subsubsection{How Hyperparameters Affect Results}
While deep learning models automatically adjust their internal parameters, their hyperparameters still require fine-tuning.
These hyperparameters can have major impact on model performance and robustness.
Some visual analytics systems expose model hyperparameters to the user for interactive experimentation.
One example previously mentioned is TensorFlow Playground~\cite{smilkov2017direct}, where users can use direct manipulation to adjust the architecture of a simple, fully-connected neural network, as well as the hyperparameters associated with its training, such as the learning rate, activation function, and regularization.
Another example is a Distill article that meticulously explores the hyperparaemters of the t-SNE dimensionality reduction method~\cite{wattenberg2016how}.
This article tests dozens of synthetic datasets in different arrangements, while varying hyperparameters such as the t-SNE perplexity and the number of iterations to run the algorithm for.

\subsection{Algorithms for Attribution \& Feature Visualization}
\label{subsec:how-algorithms}
The final method for how to visualize deep learning hails from the AI and computer vision communities.
These are algorithmic techniques that entail image generation.
Given a trained a model, one can select a single image instance and use one of the algorithmic techniques to generate a new image of the same size that either highlights important regions of the image (often called \textit{attribution}) or is an entirely new image that supposedly is representative of the same class (often called \textit{feature visualization})~\cite{olah2017feature, olah2018the}.
In these works, it is common to see large, full-page figures consisting of hundreds of such images corresponding to multiple images classes~\cite{mahendran2016visualizing}.
However, it is uncommon to see interactivity in these works, as the primary contribution is often about algorithms, not interactive techniques or systems.
Since the focus of this interrogative survey is on visual analytics in deep learning, we do not discuss in detail the various types of algorithmic techniques.
Rather, we mention the most prominent techniques developed, since they are  impactful to the growing field of deep learning visualization and could be incorporated into visual analytics systems in the future.
For more details about these techniques, such as input modification, deconvolutional methods~\cite{zeiler2014visualizing}, and input reconstruction methods, 
we refer our readers to the taxonomies~\cite{grun2016taxonomy} and literature surveys for visualizing learned features in CNNs~\cite{seifert2017visualizations, kindermans2017patternnet},
and a tutorial that presents the theory behind many of these interpretation techniques and discusses tricks and recommendations to efficiently use them on real data~\cite{montavon2017methods}.

\subsubsection{Heatmaps for Attribution, Attention, \& Saliency}
One research area generates translucent heatmaps that overlay images to highlight important regions that contribute towards classification and their sensitivity~\cite{simonyan2013deep, li2017beyond, smilkov2017smoothgrad, selvaraju2016grad, zhou2016learning}.
One technique called visual backpropagation attempts to visualize which parts of an image have contributed to the classification, and can do so in real-time in a model debugging tool for self-driving vehicles~\cite{bojarski2016visualbackprop}.
Another technique is to invert representations, i.e., attempt to reconstruct an image using a feature vector to understand the what a CNN has learned~\cite{dosovitskiy2016inverting, mahendran2015understanding, yu2014visualizing}.
Prediction difference analysis is a method that highlights features in an image to provide evidence for or against a certain class\cite{zintgraf2017visualizing}.
Other work hearkens back to more traditional computer vision techniques by exploring how object detectors emerge in CNNs and attempts to give humans object detector vision capabilities to better align humans and deep learning vision for images~\cite{zhou2014object, vondrick2013hoggles}.
Visualizing CNN filters is also popular, and has famously shown to generate dream-like images, becoming popular in artistic tasks~\cite{mordvintsev2015inceptionism, springenberg2014striving} .
Some work for interpreting \textit{visual question answering (VQA)} models and tasks use these heatmaps to explain which parts of an image a VQA model is looking at in unison with text activation maps when answering the given textual questions~\cite{goyal2016towards}.
However, recent work has shown that some of these methods fail to provide correct results and argue that we should develop explanation methods that work on simpler models before extending them to the more complex ones~\cite{kindermans2017patternnet}. 
\subsubsection{Feature Visualization}
For feature visualization, while some techniques have proven interesting~\cite{wei2015understanding}, one of the most studied techniques, class activation maximization, maximizes the activation of a chosen, specific neuron using an optimization scheme, such as gradient ascent, and generates synthetic images that are representative of what the model has learned about the chosen class~\cite{erhan2009visualizing}.
This led to a number of works improving the quality of the generated images.
Some studies generated hundreds of these non-deterministic synthetic images and clustered them to see how variations in the class activation maximization algorithm affects the output image~\cite{nguyen2016multifaceted}.
In some of their most recent work on this topic, Ngyuen et al.~\cite{nguyen2016synthesizing} present hundreds of high-quality images using a deep generator network to improve upon the state-of-the-art, and include figures comparing their technique to many of the existing and previous attempts to improve the quality of generated images.
The techniques developed in this research area have improved dramatically over the past few years, where now it is possibly to synthetically generate photorealistic  images~\cite{nguyen2016plug}.
A recent comparison of feature visualization techniques highlights their usefulness~\cite{olah2017feature}; however, the authors note that they remain skeptical of their trustworthiness, e.g., do neurons have a consistent meaning across different inputs, and if so, is that meaning accurately reified by feature visualization~\cite{olah2018the}?
\section{When to Visualize in the Deep Learning Process}
\label{sec:when}

This section describes when visualizing deep learning may be most relevant and useful.
Our discussion primarily centers around the training process: an iterative, foundational procedure for using deep learning models.
We identify two distinct, non-mutually exclusive times for when to visualize: \textit{during training} and \textit{after training}.
Some works propose that visualization be used both during and after training.

\subsection{During Training}
\label{subsec:when-during}
Artificial neural networks learn higher-level features that are useful for class discrimination as training progress~\cite{bengio2009learning}.
By using visualization during the training process, there is potential to monitor one's model as it learns to closely observe and track the model's performance~\cite{rauber2017visualizing}.

Many of the systems in this category run in a separate web-browser alongside the training process, and interface with the underlying model to query the latest model status.
This way, users can visually explore and rigorously monitor their models in real time, while they are trained elsewhere.
The visualization systems dynamically update the charts with metrics recomputed after every epoch, e.g., the loss, accuracy, and training time.
Such metrics are important to model developers because they rely on them to determine if a model
(1) has begun to learn anything at all; 
(2) is converging and reaching the peak of its performance; or 
(3) has potentially overfitted and memorized the training data.
Therefore, many of the visual analytics systems used during training support and show these updating visualizations as a primary view in the interface~\cite{abadi2016tensorflow, smilkov2017direct, chung2016revacnn, pezzotti2017deepeyes, liu2018analyzing, chae2017visualization}.
One system, Deep View~\cite{zhong2017evolutionary}, visualizes model metrics during the training process and uses its own defined metrics for monitoring (rather than the loss): a discriminability metric, which evaluates neuron evolution, and a density metric which evaluates the output feature maps.
This way, for detecting overfitting, the user does not need to wait long to view to infer overfitting; they simply observe the neuron density early in training phase.

Similarly, some systems help reduce development time and save computational resources by visualizing metrics that indicate whether a model is successfully learning or not, allowing a user to stop the training process early~\cite{smilkov2017direct}.
By using visualization during model training, users can save development time through model steering~\cite{chung2016revacnn} and utilizing suggestions for model improvement~\cite{chae2017visualization}.
Lastly, another model development time minimization focuses on diagnosing neurons and layers that are not training correctly or are misclassifying data instances.
Examples include 
DeepEyes~\cite{pezzotti2017deepeyes}, a system that identifies stable and unstable layers and neurons so users may prune their models to speed up training; 
\textit{Blocks}~\cite{bilal2018convolutional}, a system that visualizes class confusion and reveals that confusion patterns follow a hierarchical structure over the classes which can then be exploited to design hierarchy-aware architectures; 
and DGMTracker~\cite{liu2018analyzing}, a system that proposes a credit assignment algorithm that indicates how other neurons contribute to the output of particular failing neurons.

\subsection{After Training}
\label{subsec:when-after}
While some works support neural network design during the iterative model building process, there are other works that focus their visualization efforts after a model has been trained.
In other words, these works assume a trained model as input to the system or visualization technique.
Note that many, if not most, of the previously mentioned algorithmic techniques developed in the AI fields, such as attribution and feature visualization, are performed after training.
These techniques are discussed more in Section \ref{subsec:how-algorithms}.

The Embedding Projector~\cite{smilkov2016embedding} specializes in visualizing 2D and 3D embeddings produced by trained neural networks.
While users can visualize typical high-dimensional datasets in this tool, the Embedding Projector tailors the experience towards embeddings commonly used deep learning.
Once a neural network model has been trained, one can compute the activations for a given test dataset and visualize the activations in the Embedding Projector to visualize and explore the space that the network has learned.
Instead of generating an overview embedding, another previously discussed system, the Deep Visualization Toolbox~\cite{yosinski2015understanding}, uses a trained model to visualize live activations in a large small-multiples view to understand of what types of filters a convolutional network has learned.

More traditional visual analytics systems have also been developed to inspect a model after it has finished training.
ActiVis~\cite{kahng2018activis}, a visual analytics system for neural network interpretation deployed at Facebook reports that Facebook engineers and data scientists use visual analytics systems often in their normal workflow.
Another system, RNNVis~\cite{ming2017understanding}, visualizes and compares different RNN models for various natural language processing tasks.
This system positions itself as a natural extension of TensorFlow; using multiple TensorFlow models as input, the system then analyzes the trained models to extract learned representations in hidden states, and further processes the evaluation results for visualization.
Lastly, the LSTMVis~\cite{strobelt2017lstmvis} system, a visual analysis tool for RNN interpretability, separates model training from the visualization.
This system takes a model as input that must be trained separately, and from the model, gathers the required information to produce the interactive visualizations to be rendered in a web-based front-end.

\section{Where is Deep Learning Visualization}
\label{sec:where}

For the last question of the interrogative survey, we divide up ``Where'' into two subsections: where deep learning visualization research has been applied, and where deep learning visualization research has been conducted, describing the new and hybrid community.
This division provides a concise summary for practitioners who wish to investigate the usage of the described techniques for their own work, and provides new researchers with the main venues for this research area to investigate existing literature.

\subsection{Application Domains \& Models}
\label{subsec:where-application}
While many non-neural approaches are used for real-world applications, deep learning has successfully achieved state-of-the-art performance in several domains.
Previously in Section \ref{subsubsec:application}, we presented works that apply neural networks to particular domains and use visualizations to lend qualitative support to their usual quantitative results to strengthen users' trust in their models.
These domains included neural machine translation~\cite{johnson2016google}, reinforcement learning~\cite{zahavy2016graying}, social good~\cite{robinson2017deeppop}, autonomous vehicles~\cite{bojarski2016visualbackprop}, medical imaging diagnostics~\cite{zintgraf2017visualizing}, and urban planning~\cite{li2017hierarchical}.

Next we summarize the types of models that have been used in deep learning visualization.
Much of the existing work has used image-based data and models, namely CNNs, to generate attribution and feature visualization explanations for what a model has learned from an image dataset.
CNNs, while not exclusively used for images, have become popular in the computer vision community and are often used for image classification and interactive, image-based creative tasks~\cite{zhu2016generative, carter2017using}.
Besides images, sequential data (e.g., text, time series data, and music) has also been studied.
This research stems from the natural language processing community, where researchers typically favor RNNs for learning representations of large text corpora.
These researchers make sense of large word embeddings by using interactive tools that support dimensionality reduction techniques to solve problems such as sequence-to-sequence conversion, translation, and audio recognition.
Research combining both image and text data has also been done, such as image captioning and visual question answering~\cite{vinyals2015show, antol2015vqa}.
Harder still are new types of networks called \textit{generative adversarial networks}, or GANs for short, that have produced remarkable results for data generation~\cite{goodfellow2014generative}, e.g., producing real-looking yet fake images~\cite{goodfellow2016nips}.
While GANs have only existed for a couple of years, they are now receiving significant research attention.
To make sense of the learned features and distributions from GANs, two visual analytics systems, DGMTracker~\cite{liu2018analyzing} and GANViz~\cite{wang2018ganviz}, focus on understanding the training dynamics of GANs to help model developers better train these complex models, often consisting of multiple dueling neural networks.

\subsection{A Vibrant Research Community: Hybrid, Apace, \& Open-sourced}
\label{subsec:where-community}
As seen from this survey, bringing together the visualization communities with the AI communities has led to the design and development of numerous tools and techniques for improving deep learning interpretability and democratization.
This hybrid research area has seen accelerated attention and interest due to its widespread impact, as evidenced by the large number of works published in just a few years, as seen in Table \ref{table:papers}.
A consequence of this rapid progress is that deep learning visualization research are being disseminated across multiple related venues.
In academia, the premiere venues for deep learning visualization research consists of two main groups: the information visualization and visual analytics communities; and the artificial intelligence and deep learning communities.
Furthermore, since this area is relatively new, it has seen more attention at multiple workshops at the previously mentioned academic conferences, as tabulated in Table \ref{table:venues}.

Another consequence of this rapidly developing area is that new work is immediately publicized and open-sourced, without waiting for it to be ``officially'' published at conferences, journals, etc.
Many of these releases take the form of a preprint publication posted on arXiv, where a deep learning presence has thrived.
Not only is it common for academic research labs and individuals to publish work on arXiv, but companies from industry are also publishing results, code, and tools.
For example, the most popular libraries\footnote{Popular libraries include
TensorFlow~\cite{abadi2016tensorflow},
Keras,
Caffe,
PyTorch,
and Theano.
}
for implementing neural networks are open-source and have consistent contributions for improving all areas of the codebase, e.g., installation, computation, and deployment into specific programming languages' open-source environments.

Some works have a corresponding blog post on an industry research blog\footnote{High impact industry blogs include: Google Research Blog, OpenAI, Facebook Research Blog, the Apple Machine Learning Journal, NVIDIA Deep Learning AI, and Uber AI}, which, while non-traditional, has large impact due to their prominent visibility and large readership.
While posting preprints may have its downsides (e.g., little quality control) 
the communities have been promoting the good practices of open-sourcing developed code and including direct links within the preprints; both practices are now the norm.
Although it may be overwhelming to digest the amount of new research published daily, having access to the work with its code could encourage reproducibility and allow the communities to progress faster.
In summary, given the increasing interest in deep learning visualization research and its importance, we believe our communities will continue to thrive, and will positively impact many domains for years to come.

\section{Research Directions \& Open Problems}
\label{sec:open-challenges}

Now we present research directions and open problems for future research distilled from the surveyed works.

\subsection{Furthering Interpretability}
Unsurprisingly, with the amount of attention and importance on interpretability and explainability in the deep learning visualization literature, the first area for future work is continuing to create new interpretable methods for deep learning models.
For the information visualization and visual analytics communities, this could constitute creating new visual representations for the components in deep learning models, or developing new interaction techniques to reveal deeper insights about one's model.
For the AI communities, more insightful attribution and feature visualization techniques for trained models that are fast (computationally cheap) could be incorporated into visualization systems.
Combining visual representations, helpful interactions, and state-of-the-art attribution and feature visualization techniques together into rich user interfaces could lead to major breakthroughs for understanding neural networks~\cite{olah2018the}.

\begin{figure}[tb]
 \centering
 \includegraphics[width=\columnwidth]{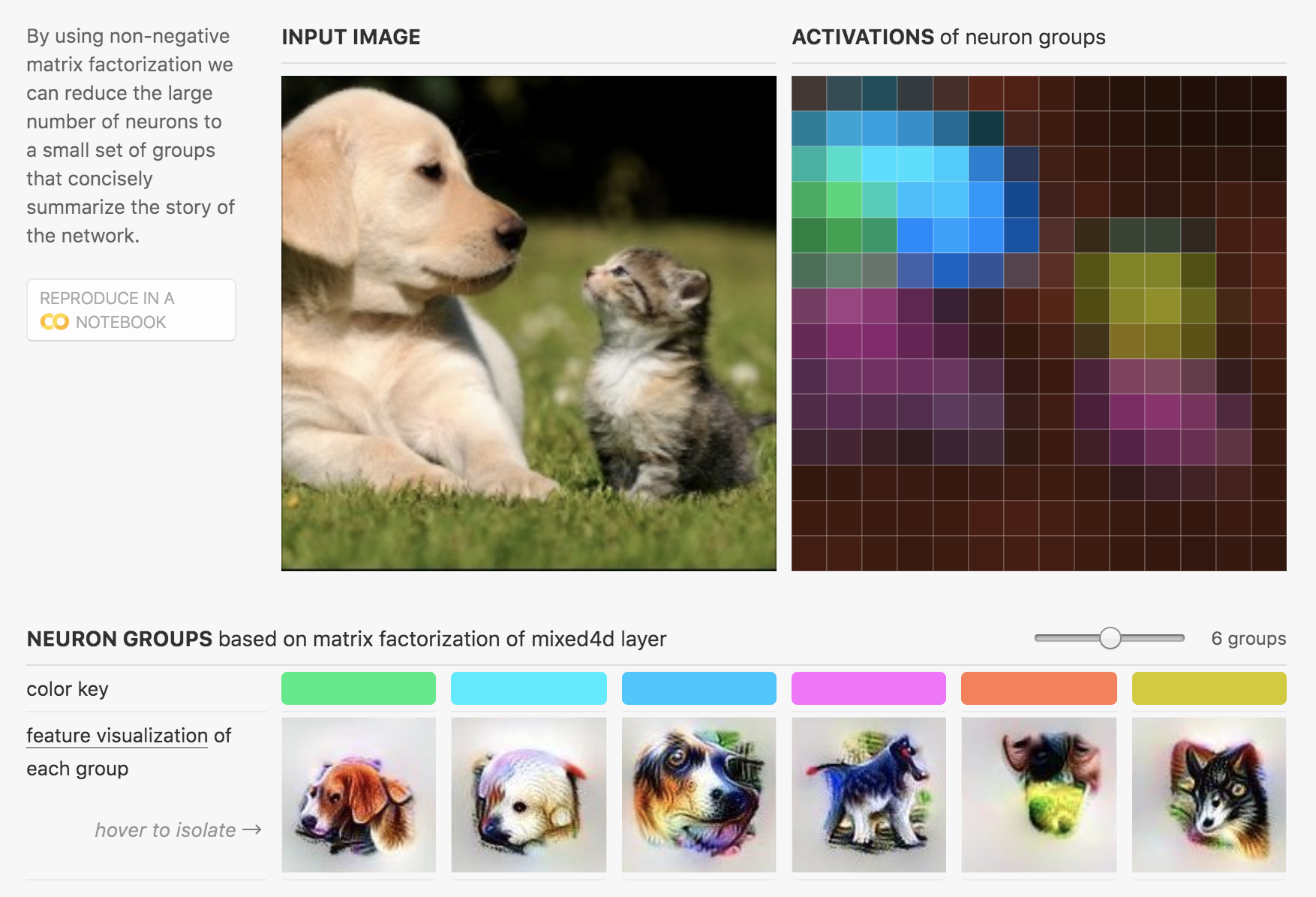}
 \caption{
 Distill: The Building Blocks of Interpretability~\cite{olah2018the}: an interactive user interface that combines feature visualization and attribution techniques to interpret neural networks.
 }
 \label{fig:playground}
\end{figure}

\subsection{System \& Visual Scalability}
Throughout this survey, we have covered many visual analytics systems that  facilitate interpretation and model understanding.
However, some systems suffer from scalability problems.
Visual scalability challenges arise when dealing with large data, e.g., large number of hyperparameters and millions of parameters in deep neural networks.
Some research has started to address this, by simplifying complex dataflow graphs and network weights for better model explanations~\cite{wongsuphasawat2018visualizing, liu2017towards, strobelt2017lstmvis}.
But, regarding activations and embeddings, dimensionality reduction techniques have a limit to their usability when it comes to the number of points to visualize~\cite{rauber2017visualizing}.
We think this is an important research direction, especially given that the information visualization communities have developed techniques to visualize large, high-dimensional data that could  potentially be applicable to deep learning~\cite{liu2017visualizing}.

Aside from visual scalability, some tools also suffer from system scalability.
While some of these problems may be more engineering-centric, we think that for visual analytics systems to adopted, they need to handle state-of-the-art deep models without penalizing performance or increasing model development time.
Furthermore, these systems (often web-based) will greatly benefit from fast computations, supporting real-time, rich user interactions~\cite{rong2016visual}.
This is especially important for visual systems that need to perform pre-computation before rendering visualizations to the screen.

\subsection{Design Studies for Evaluation: Utility \& Usability}
An important facet of visualization research is the evaluation of the utility and usefulness of the visual representation. 
Equally important is to evaluate the usability of deployed systems and their interactive visual analytics techniques.
It is encouraging to see many of the visual analytics systems recognize this importance and report on design studies conducted with AI experts before building a tool to understand the users and their needs~\cite{wongsuphasawat2018visualizing, kahng2018activis, strobelt2017lstmvis, liu2017towards, ming2017understanding, ren2017squares}.
It is common to see example use cases or illustrative usage scenarios that demonstrate the capabilities of the interactive systems. 
Some works go beyond these and conduct user studies to evaluate utility and usability~\cite{bruckner2014mloscope}.
In the AI communities, most works do not include user studies.
For those that do, they greatly benefit from showing why their proposed methods are superior to the ones being tested against~\cite{selvaraju2016grad, samek2017evaluating, mahendran2016visualizing, ribeiro2016should}.
Taking this idea to the quantifiable extreme, a related avenue of evaluating these techniques is the notion of quantifying interpretability, which has been recently studied~\cite{bau2017netdissect, tsa2016icharacterizing}.
Other domains have recognized that interpretable deep learning research may require evaluation techniques for their interpretations, and argue that there is a large body of work from fields such as philosophy, cognitive science, and social psychology that could be utilized~\cite{miller2017explanation, ritter2017cognitive}.

When surveying the interfaces of deep learning visual analytics tools, many of them contain multiple-coordinated views with many visual representations.
Displaying this much information at once can be overwhelming, and when interpretability is the primary focus, it is critical for these systems to have superior usability.
Therefore, we think future works could further benefit from including more members of the human-computer interaction communities, including interface and user experience designers, that could help organize and prioritize interfaces using well-studied guidelines~\cite{weld2018intelligible}.

\subsection{The Human Role in Interpretability}
\subsubsection{Human v. Machine Understanding of the World}
In deep learning interpretability work, researchers are developing methods to produce better explanations to ``see through the black-box,'' but unfortunately some of these methods produce visualizations that, while visually interesting and thought-provoking~\cite{mordvintsev2015inceptionism}, are not fully understandable by their human viewers.
That is an important facet of deep learning interpretability, namely, producing visualizations and interpretations that are human understandable~\cite{olah2017feature}.
Some methods compare algorithmic results with an empirically derived human baseline; this enables comparison between machine and human generated responses to objects in the world, particularly in images~\cite{das2017human}.
Ultimately, researchers seek to understand the commonalities and differences between how humans and machines see and decompose the world~\cite{tam2017analysis}.
Some tools that we have surveyed achieve this by using live-video to compare the input images and the neural network's activations and filters in real time~\cite{yosinski2015understanding}.
Other tools give users explicit control of an experiment by training multiple small models with only a few exposed hyperparameters, automatically generating visualizations to then see the effect that the input data has on the learned representation~\cite{hohman2017shapeshop}.
These ``what-if'' tools and scenarios could potentially be extended to incorporate human feedback into the training or model steering process of neural network to better improve performance.

\subsubsection{Human-AI Pairing}
Much of this survey is dedicated towards reviewing the state-of-the-art in visual analytics for deep learning, with a focus on interpretability.
These works use visualization to explain, explore, and debug models in order to choose the best preforming model for a given task, often by placing a human in the loop.
However, a slight twist on this idea hearkening back to the original envisioning of the computer has lead to the emergence of a new research area, one where tasks are not exclusively performed by humans or machines, but one where the two complement each other.
This area, recently dubbed \textit{artificial intelligence augmentation} describes the use of AI systems to help develop new methods for intelligence augmentation~\cite{carter2017using}.
Some related works we have covered already propose artificial intelligence augmentation ideas, such as a system that suggests potentially interesting directions to explore in a high-dimensional 3D embedding~\cite{smilkov2016embedding}, predicting and showing where the next stroke of a word could be when handwriting text~\cite{carter2016experiments}, automatically generating images based off of user-provided sketches~\cite{zhu2016generative}, and dynamically changing and steering a neural network model while it trains~\cite{chung2016revacnn}.
We believe this is a rich, under-explored area for future research: using well-designed interfaces for humans to interact with machine learning models, and for these machine learning models to augment creative human tasks.

\subsection{Social Good \& Bias Detection}
The aspirational pairing of humans and machines is a long-term research endeavor.
To quicken our pace, we must continue to democratize artificial intelligence via educational tools, perhaps by using direct manipulation as an invitation for people to engage with AI~\cite{smilkov2017direct, olah2018the}, clear explanations for model decision making, and robust tooling and libraries for programming languages for people to develop such models~\cite{abadi2016tensorflow, wongsuphasawat2018visualizing}.
While doing this, we must also ensure that AI applications remain ethical, fair, safe, transparent, and are benefiting society~\cite{weller2017challenges}.

Another important consideration for future research is detecting bias.
This has been identified as a major problem in deep learning \cite{barocas2016big, caliskan2017semantics}, and a number of researchers are using visualization to understand why a model may be biased~\cite{wattenberg2016attacking}.
One example that aims to detect data bias is Google's Facets tool~\cite{2017facets}, a visual analytics system designed specifically to preview and visualize machine learning datasets before training.
This allows one to inspect large datasets by exploring the different classes or data instances,  to see if there are any high-level imbalances in the class or data distribution.

Other works have begun to explore if the mathematical algorithms themselves can be biased towards particular decisions. 
An example of this is an interactive article titled ``Attacking discrimination with smarter machine learning''~\cite{wattenberg2016attacking}, which explores how one can can create both fair and unfair threshold classifiers in an example task such as loan granting scenarios where a bank may grant or deny a loan based on a single, automatically computed number such as a credit score.
The article aims to highlight that \textit{equal opportunity}~\cite{hardt2016equality} is not preserved by machine learning algorithms, and that as AI-powered systems continue to make important decisions across core social domains, it is critical to ensure decisions are not discriminatory.

Finally, aside from data and model bias, humans are often inherently biased decision makers. 
In response, there is a growing area of research into detecting and understanding bias in visual analytics
\footnote{The DECISIVe Workshop (\url{http://decisive-workshop.dbvis.de/}) at IEEE VIS is dedicated to understanding cognitive bias in visualization.}
and its affect on the decision making process~\cite{wall2017warning}.
Some work has developed metrics to detect types of bias to present to a user during data analysis~\cite{wall2017warning} which could also be applied to visual tools for deep learning in the future.
Some work has employed developmental and cognitive psychology analysis techniques to understand how humans learn, focusing on uncovering how human bias is developed and influences learning, to ultimately influence artificial neural network design~\cite{ritter2017cognitive}.

\subsection{Protecting Against Adversarial Attacks}
Regardless of the benefits AI systems are bringing to society, we would be remiss to immediately trust them; like most technologies, AI too has security faults.
Identified and studied in seminal works, it has been shown that deep learning models such as image classifiers can be easily fooled by perturbing an input image~\cite{szegedy2013intriguing, goodfellow2014explaining, nguyen2015deep}.
Most alarming, some perturbations are so subtle that they are untraceable by the human eye, yet would completely fool a model into misclassification~\cite{goodfellow2014explaining}.
This sparked great interest in the AI communities, and much work has been done to understand how fragile deep neural network image classifiers are, identify in what ways can they break, and explore methods for protecting them.
Norton et al.~\cite{norton2017adversarial} demonstrate adding adversarial perturbations to images in an interactive tool, where users can tweak the type and intensity of the attack, and observe the resulting (mis)classification.
This is a great first start for using visualization to identify potential attacks, but we think visualization can be majorly impactful in this research space, by not only showcasing how the attacks work and detecting them, but also by taking action and protecting AI systems from the attacks themselves.
While some work, primarily originating from the AI communities, has proposed computational techniques to protect AI from attacks, such as identifying adversarial examples before classification~\cite{metzen2017detecting}, modifying the network architecture~\cite{gu2014towards}, modifying the training process~\cite{papernot2016distillation, goodfellow2014explaining}, and performing pre-processing steps before classification~\cite{das2018shield, kurakin2016adversarial}, we think visualization can have great impact for combating adversarial machine learning.

\section{Conclusion}
\label{sec:conclusion}

We presented a comprehensive, timely survey on  visualization and visual analytics in deep learning research, using a human-centered, interrogative framework.
Our method helps researchers and practitioners in visual analytics and deep learning to quickly learn key aspects of this young and rapidly growing body of research, whose impact spans a broad range of domains.
Our survey goes beyond visualization-focused venues to extend a wide scope that also encompasses relevant works from top venues in AI, ML, and computer vision.
We highlighted  visual analytics as an integral component in addressing pressing issues in modern AI, helping to discover and communicate insight, from discerning model bias, understanding models, to promoting AI safety.
We concluded by highlighting impactful research directions and open problems.


%



\ifCLASSOPTIONcompsoc
  \section*{Acknowledgments}
\else
  \section*{Acknowledgment}
\fi

This work was supported by NSF grants IIS-1563816, CNS-1704701, and TWC-1526254; NIBIB grant U54EB020404; NSF GRFP DGE-1650044; NASA Space Technology Research Fellowship; and gifts from Intel, Google, Symantec.

\ifCLASSOPTIONcaptionsoff
  \newpage
\fi



%
\bibliographystyle{IEEEtran}
\bibliography{IEEEabrv,main}

\begin{thebibliography}{100}
\providecommand{\url}[1]{#1}
\csname url@samestyle\endcsname
\providecommand{\newblock}{\relax}
\providecommand{\bibinfo}[2]{#2}
\providecommand{\BIBentrySTDinterwordspacing}{\spaceskip=0pt\relax}
\providecommand{\BIBentryALTinterwordstretchfactor}{4}
\providecommand{\BIBentryALTinterwordspacing}{\spaceskip=\fontdimen2\font plus
\BIBentryALTinterwordstretchfactor\fontdimen3\font minus
  \fontdimen4\font\relax}
\providecommand{\BIBforeignlanguage}[2]{{%
\expandafter\ifx\csname l@#1\endcsname\relax
\typeout{** WARNING: IEEEtran.bst: No hyphenation pattern has been}%
\typeout{** loaded for the language `#1'. Using the pattern for}%
\typeout{** the default language instead.}%
\else
\language=\csname l@#1\endcsname
\fi
#2}}
\providecommand{\BIBdecl}{\relax}
\BIBdecl

\bibitem{mcculloch1943logical}
W.~S. McCulloch and W.~Pitts, ``A logical calculus of the ideas immanent in
  nervous activity,'' \emph{The bulletin of mathematical biophysics}, vol.~5,
  no.~4, 1943.

\bibitem{rawat2017deep}
W.~Rawat and Z.~Wang, ``Deep convolutional neural networks for image
  classification: A comprehensive review,'' \emph{Neural computation}, vol.~29,
  no.~9, 2017.

\bibitem{krizhevsky2012imagenet}
A.~Krizhevsky, I.~Sutskever, and G.~E. Hinton, ``{ImageNet} classification with
  deep convolutional neural networks,'' in \emph{NIPS}, 2012.

\bibitem{simonyan2013deep}
K.~Simonyan, A.~Vedaldi, and A.~Zisserman, ``Deep inside convolutional
  networks: Visualising image classification models and saliency maps,''
  \emph{arXiv:1312.6034}, 2013.

\bibitem{szegedy2015going}
C.~Szegedy, W.~Liu, Y.~Jia, P.~Sermanet, S.~Reed, D.~Anguelov, D.~Erhan,
  V.~Vanhoucke, and A.~Rabinovich, ``Going deeper with convolutions,'' in
  \emph{CVPR}, 2015.

\bibitem{karpathy2014learned}
\BIBentryALTinterwordspacing
A.~Karpathy, ``What {I} learned from competing against a convnet on
  {ImageNet},'' 2014. [Online]. Available: \url{http://karpathy.github.
  io/2014/09/02/what-i-learned-from-eompeting-against-a-convnet-on-imagenet}
\BIBentrySTDinterwordspacing

\bibitem{he2016deep}
K.~He, X.~Zhang, S.~Ren, and J.~Sun, ``Deep residual learning for image
  recognition,'' in \emph{CVPR}, 2016.

\bibitem{deng2009imagenet}
J.~Deng, W.~Dong, R.~Socher, L.-J. Li, K.~Li, and L.~Fei-Fei, ``{ImageNet}: A
  large-scale hierarchical image database,'' in \emph{CVPR}, 2009.

\bibitem{russakovsky2015imagenet}
O.~Russakovsky, J.~Deng, H.~Su, J.~Krause, S.~Satheesh, S.~Ma, Z.~Huang,
  A.~Karpathy, A.~Khosla, M.~Bernstein \emph{et~al.}, ``Imagenet large scale
  visual recognition challenge,'' \emph{IJCV}, vol. 115, no.~3, 2015.

\bibitem{zeiler2014visualizing}
M.~D. Zeiler and R.~Fergus, ``Visualizing and understanding convolutional
  networks,'' in \emph{ECCV}.\hskip 1em plus 0.5em minus 0.4em\relax Springer,
  2014.

\bibitem{craven1992visualizing}
M.~W. Craven and J.~W. Shavlik, ``Visualizing learning and computation in
  artificial neural networks,'' \emph{International Journal on Artificial
  Intelligence Tools}, vol.~1, no.~03, 1992.

\bibitem{streeter2001nvis}
M.~J. Streeter, M.~O. Ward, and S.~A. Alvarez, ``{NVIS}: An interactive
  visualization tool for neural networks,'' in \emph{Visual Data Exploration
  and Analysis VIII}, vol. 4302.\hskip 1em plus 0.5em minus 0.4em\relax
  International Society for Optics and Photonics, 2001.

\bibitem{tzeng2005opening}
F.-Y. Tzeng and K.-L. Ma, ``Opening the black box: Data driven visualization of
  neural networks,'' in \emph{IEEE Visualization}, 2005.

\bibitem{liu2017towards}
M.~Liu, J.~Shi, Z.~Li, C.~Li, J.~Zhu, and S.~Liu, ``Towards better analysis of
  deep convolutional neural networks,'' \emph{IEEE TVCG}, vol.~23, no.~1, 2017.

\bibitem{wongsuphasawat2018visualizing}
K.~Wongsuphasawat, D.~Smilkov, J.~Wexler, J.~Wilson, D.~Man{\'e}, D.~Fritz,
  D.~Krishnan, F.~B. Vi{\'e}gas, and M.~Wattenberg, ``Visualizing dataflow
  graphs of deep learning models in {TensorFlow},'' \emph{IEEE TVCG}, vol.~24,
  no.~1, 2018.

\bibitem{smilkov2017direct}
D.~Smilkov, S.~Carter, D.~Sculley, F.~B. Viegas, and M.~Wattenberg,
  ``Direct-manipulation visualization of deep networks,'' in \emph{ICML
  Workshop on Vis for Deep Learning}, 2016.

\bibitem{lu2017recent}
J.~Lu, W.~Chen, Y.~Ma, J.~Ke, Z.~Li, F.~Zhang, and R.~Maciejewski, ``Recent
  progress and trends in predictive visual analytics,'' \emph{Frontiers of
  Computer Science}, 2017.

\bibitem{lu2017state}
Y.~Lu, R.~Garcia, B.~Hansen, M.~Gleicher, and R.~Maciejewski, ``The
  state-of-the-art in predictive visual analytics,'' in \emph{Computer Graphics
  Forum}, vol.~36, no.~3.\hskip 1em plus 0.5em minus 0.4em\relax Wiley Online
  Library, 2017.

\bibitem{ren2017squares}
D.~Ren, S.~Amershi, B.~Lee, J.~Suh, and J.~D. Williams, ``Squares: Supporting
  interactive performance analysis for multiclass classifiers,'' \emph{IEEE
  TVCG}, vol.~23, no.~1, 2017.

\bibitem{amershi2014power}
S.~Amershi, M.~Cakmak, W.~B. Knox, and T.~Kulesza, ``Power to the people: The
  role of humans in interactive machine learning,'' \emph{AI Magazine},
  vol.~35, no.~4, 2014.

\bibitem{sacha2016human}
D.~Sacha, M.~Sedlmair, L.~Zhang, J.~A. Lee, D.~Weiskopf, S.~North, and D.~Keim,
  ``Human-centered machine learning through interactive visualization,'' in
  \emph{ESANN}, 2016.

\bibitem{seifert2017visualizations}
C.~Seifert, A.~Aamir, A.~Balagopalan, D.~Jain, A.~Sharma, S.~Grottel, and
  S.~Gumhold, ``Visualizations of deep neural networks in computer vision: A
  survey,'' in \emph{Transparent Data Mining for Big and Small Data}.\hskip 1em
  plus 0.5em minus 0.4em\relax Springer, 2017.

\bibitem{zeng2017towards}
H.~Zeng, ``Towards better understanding of deep learning with visualization,''
  \emph{The Hong Kong University of Science and Technology}, 2016.

\bibitem{liu2017towardsva}
S.~Liu, X.~Wang, M.~Liu, and J.~Zhu, ``Towards better analysis of machine
  learning models: A visual analytics perspective,'' \emph{Visual Informatics},
  vol.~1, no.~1, 2017.

\bibitem{choo2018visual}
J.~Choo and S.~Liu, ``Visual analytics for explainable deep learning,''
  \emph{IEEE Computer Graphics and Applications}, 2018.

\bibitem{Goodfellow-et-al-2016}
I.~Goodfellow, Y.~Bengio, and A.~Courville, \emph{Deep Learning}.\hskip 1em
  plus 0.5em minus 0.4em\relax MIT Press, 2016,
  \url{http://www.deeplearningbook.org}.

\bibitem{abadi2016tensorflow}
M.~Abadi, A.~Agarwal, P.~Barham, E.~Brevdo, Z.~Chen, C.~Citro, G.~S. Corrado,
  A.~Davis, J.~Dean, M.~Devin \emph{et~al.}, ``{TensorFlow}: Large-scale
  machine learning on heterogeneous distributed systems,''
  \emph{arXiv:1603.04467}, 2016.

\bibitem{bau2017netdissect}
D.~Bau, B.~Zhou, A.~Khosla, A.~Oliva, and A.~Torralba, ``{Network Dissection}:
  Quantifying interpretability of deep visual representations,'' in
  \emph{CVPR}, 2017.

\bibitem{bilal2018convolutional}
A.~Bilal, A.~Jourabloo, M.~Ye, X.~Liu, and L.~Ren, ``Do convolutional neural
  networks learn class hierarchy?'' \emph{IEEE TVCG}, vol.~24, no.~1, pp.
  152--162, 2018.

\bibitem{bojarski2016visualbackprop}
M.~Bojarski, A.~Choromanska, K.~Choromanski, B.~Firner, L.~Jackel, U.~Muller,
  and K.~Zieba, ``Visualbackprop: visualizing cnns for autonomous driving,''
  \emph{arXiv:1611.05418}, 2016.

\bibitem{bruckner2014mloscope}
\BIBentryALTinterwordspacing
D.~Bruckner, ``Ml-o-scope: a diagnostic visualization system for deep machine
  learning pipelines,'' Master's thesis, EECS Department, University of
  California, Berkeley, May 2014. [Online]. Available:
  \url{http://www2.eecs.berkeley.edu/Pubs/TechRpts/2014/EECS-2014-99.html}
\BIBentrySTDinterwordspacing

\bibitem{carter2016experiments}
S.~Carter, D.~Ha, I.~Johnson, and C.~Olah, ``Experiments in handwriting with a
  neural network,'' \emph{Distill}, 2016.

\bibitem{cashman2017rnnbow}
D.~Cashman, G.~Patterson, A.~Mosca, and R.~Chang, ``{RNNbow}: Visualizing
  learning via backpropagation gradients in recurrent neural networks,'' in
  \emph{Workshop on Visual Analytics for Deep Learning}, 2017.

\bibitem{chae2017visualization}
J.~Chae, S.~Gao, A.~Ramanthan, C.~Steed, and G.~D. Tourassi, ``Visualization
  for classification in deep neural networks,'' in \emph{Workshop on Visual
  Analytics for Deep Learning}, 2017.

\bibitem{chung2016revacnn}
S.~Chung, C.~Park, S.~Suh, K.~Kang, J.~Choo, and B.~C. Kwon, ``{ReVACNN}:
  Steering convolutional neural network via real-time visual analytics,'' in
  \emph{NIPS Workshop on Future of Interactive Learning Machines}, 2016.

\bibitem{goyal2016towards}
Y.~Goyal, A.~Mohapatra, D.~Parikh, and D.~Batra, ``Towards transparent ai
  systems: Interpreting visual question answering models,''
  \emph{arXiv:1608.08974}, 2016.

\bibitem{harley2015isvc}
A.~W. Harley, ``An interactive node-link visualization of convolutional neural
  networks,'' in \emph{ISVC}, 2015, pp. 867--877.

\bibitem{hohman2017shapeshop}
F.~Hohman, N.~Hodas, and D.~H. Chau, ``{ShapeShop}: Towards understanding deep
  learning representations via interactive experimentation,'' in \emph{CHI,
  Extended Abstracts}, 2017.

\bibitem{kahng2018activis}
M.~Kahng, P.~Andrews, A.~Kalro, and D.~H. Chau, ``{ActiVis}: Visual exploration
  of industry-scale deep neural network models,'' \emph{IEEE TVCG}, vol.~24,
  no.~1, 2018.

\bibitem{karpathy2015visualizing}
A.~Karpathy, J.~Johnson, and L.~Fei-Fei, ``Visualizing and understanding
  recurrent networks,'' \emph{arXiv:1506.02078}, 2015.

\bibitem{li2015visualizing}
J.~Li, X.~Chen, E.~Hovy, and D.~Jurafsky, ``Visualizing and understanding
  neural models in nlp,'' \emph{arXiv:1506.01066}, 2015.

\bibitem{liu2018analyzing}
M.~Liu, J.~Shi, K.~Cao, J.~Zhu, and S.~Liu, ``Analyzing the training processes
  of deep generative models,'' \emph{IEEE TVCG}, vol.~24, no.~1, 2018.

\bibitem{ming2017understanding}
Y.~Ming, S.~Cao, R.~Zhang, Z.~Li, and Y.~Chen, ``Understanding hidden memories
  of recurrent neural networks,'' \emph{VAST}, 2017.

\bibitem{norton2017adversarial}
A.~P. Norton and Y.~Qi, ``{Adversarial-Playground}: A visualization suite
  showing how adversarial examples fool deep learning,'' in
  \emph{VizSec}.\hskip 1em plus 0.5em minus 0.4em\relax IEEE, 2017.

\bibitem{olah2014visualizing}
\BIBentryALTinterwordspacing
C.~Olah, ``Visualizing {MNIST},'' \emph{Olah's Blog}, 2014. [Online].
  Available: \url{http://colah.github.io/posts/2014-10-Visualizing-MNIST/}
\BIBentrySTDinterwordspacing

\bibitem{olah2018the}
C.~Olah, A.~Satyanarayan, I.~Johnson, S.~Carter, L.~Schubert, K.~Ye, and
  A.~Mordvintsev, ``The building blocks of interpretability,'' \emph{Distill},
  2018.

\bibitem{pezzotti2017deepeyes}
N.~Pezzotti, T.~H{\"o}llt, J.~Van~Gemert, B.~P. Lelieveldt, E.~Eisemann, and
  A.~Vilanova, ``{DeepEyes}: Progressive visual analytics for designing deep
  neural networks,'' \emph{IEEE TVCG}, vol.~24, no.~1, 2018.

\bibitem{rauber2017visualizing}
P.~E. Rauber, S.~G. Fadel, A.~X. Falcao, and A.~C. Telea, ``Visualizing the
  hidden activity of artificial neural networks,'' \emph{IEEE TVCG}, vol.~23,
  no.~1, 2017.

\bibitem{robinson2017deeppop}
C.~Robinson, F.~Hohman, and B.~Dilkina, ``A deep learning approach for
  population estimation from satellite imagery,'' in \emph{SIGSPATIAL Workshop
  on Geospatial Humanities}, 2017.

\bibitem{rong2016visual}
X.~Rong and E.~Adar, ``Visual tools for debugging neural language models,'' in
  \emph{ICML Workshop on Vis for Deep Learning}, 2016.

\bibitem{smilkov2016embedding}
D.~Smilkov, N.~Thorat, C.~Nicholson, E.~Reif, F.~B. Vi{\'e}gas, and
  M.~Wattenberg, ``{Embedding Projector}: Interactive visualization and
  interpretation of embeddings,'' in \emph{NIPS Workshop on Interpretable ML in
  Complex Systems}, 2016.

\bibitem{strobelt2017lstmvis}
H.~Strobelt, S.~Gehrmann, H.~Pfister, and A.~M. Rush, ``{LSTMVis}: A tool for
  visual analysis of hidden state dynamics in recurrent neural networks,''
  \emph{IEEE TVCG}, vol.~24, no.~1, 2018.

\bibitem{wang2018ganviz}
J.~Wang, L.~Gou, H.~Yang, and H.-W. Shen, ``{GANViz}: A visual analytics
  approach to understand the adversarial game,'' \emph{IEEE TVCG}, 2018.

\bibitem{webster2017teachable}
\BIBentryALTinterwordspacing
B.~Webster, ``Now anyone can explore machine learning, no coding required,''
  \emph{Google Official Blog}, 2017. [Online]. Available:
  \url{https://www.blog.google/topics/machine-learning/now-anyone-can-explore-machine-learning-no-coding-required/}
\BIBentrySTDinterwordspacing

\bibitem{yosinski2015understanding}
J.~Yosinski, J.~Clune, A.~Nguyen, T.~Fuchs, and H.~Lipson, ``Understanding
  neural networks through deep visualization,'' in \emph{ICML Deep Learning
  Workshop}, 2015.

\bibitem{zahavy2016graying}
T.~Zahavy, N.~Ben-Zrihem, and S.~Mannor, ``Graying the black box: Understanding
  {DQNs},'' in \emph{ICML}, 2016.

\bibitem{zeng2017cnncomparator}
H.~Zeng, H.~Haleem, X.~Plantaz, N.~Cao, and H.~Qu, ``{CNNComparator}:
  Comparative analytics of convolutional neural networks,'' in \emph{Workshop
  on Visual Analytics for Deep Learning}, 2017.

\bibitem{zhong2017evolutionary}
W.~Zhong, C.~Xie, Y.~Zhong, Y.~Wang, W.~Xu, S.~Cheng, and K.~Mueller,
  ``Evolutionary visual analysis of deep neural networks,'' in \emph{ICML
  Workshop on Vis for Deep Learning}, 2017.

\bibitem{zhu2016generative}
J.-Y. Zhu, P.~Kr{\"a}henb{\"u}hl, E.~Shechtman, and A.~A. Efros, ``Generative
  visual manipulation on the natural image manifold,'' in \emph{ECCV}.\hskip
  1em plus 0.5em minus 0.4em\relax Springer, 2016.

\bibitem{lipton2016mythos}
Z.~C. Lipton, ``The mythos of model interpretability,''
  \emph{arXiv:1606.03490}, 2016.

\bibitem{montavon2017methods}
G.~Montavon, W.~Samek, and K.-R. M{\"u}ller, ``Methods for interpreting and
  understanding deep neural networks,'' \emph{Digital Signal Processing}, 2017.

\bibitem{miller2017explanation}
T.~Miller, ``Explanation in artificial intelligence: Insights from the social
  sciences,'' \emph{arXiv:1706.07269}, 2017.

\bibitem{weller2017challenges}
A.~Weller, ``Challenges for transparency,'' \emph{ICML Workshop on Human
  Interpretability in ML}, 2017.

\bibitem{offert2017know}
F.~Offert, ``"i know it when {I} see it". visualization and intuitive
  interpretability,'' \emph{NIPS Symposium on Interpretable ML}, 2017.

\bibitem{johnson2016google}
M.~Johnson, M.~Schuster, Q.~V. Le, M.~Krikun, Y.~Wu, Z.~Chen, N.~Thorat,
  F.~Vi{\'e}gas, M.~Wattenberg, G.~Corrado \emph{et~al.}, ``Google's
  multilingual neural machine translation system: enabling zero-shot
  translation,'' \emph{arXiv:1611.04558}, 2016.

\bibitem{zintgraf2017visualizing}
L.~M. Zintgraf, T.~S. Cohen, T.~Adel, and M.~Welling, ``Visualizing deep neural
  network decisions: Prediction difference analysis,'' \emph{arXiv:1702.04595},
  2017.

\bibitem{li2017hierarchical}
L.~Li, J.~Tompkin, P.~Michalatos, and H.~Pfister, ``Hierarchical visual feature
  analysis for city street view datasets,'' in \emph{Workshop on Visual
  Analytics for Deep Learning}, 2017.

\bibitem{erhan2009visualizing}
D.~Erhan, Y.~Bengio, A.~Courville, and P.~Vincent, ``Visualizing higher-layer
  features of a deep network,'' \emph{University of Montreal}, vol. 1341, 2009.

\bibitem{selvaraju2016grad}
R.~R. Selvaraju, A.~Das, R.~Vedantam, M.~Cogswell, D.~Parikh, and D.~Batra,
  ``{Grad-CAM}: Why did you say that? visual explanations from deep networks
  via gradient-based localization,'' \emph{arXiv:1610.02391}, 2016.

\bibitem{nguyen2016synthesizing}
A.~Nguyen, A.~Dosovitskiy, J.~Yosinski, T.~Brox, and J.~Clune, ``Synthesizing
  the preferred inputs for neurons in neural networks via deep generator
  networks,'' in \emph{NIPS}, 2016.

\bibitem{patel2008investigating}
K.~Patel, J.~Fogarty, J.~A. Landay, and B.~Harrison, ``Investigating
  statistical machine learning as a tool for software development,'' in
  \emph{CHI}, 2008.

\bibitem{kulesza2015principles}
T.~Kulesza, M.~Burnett, W.-K. Wong, and S.~Stumpf, ``Principles of explanatory
  debugging to personalize interactive machine learning,'' in \emph{IUI}, 2015.

\bibitem{nushi2017human}
B.~Nushi, E.~Kamar, E.~Horvitz, and D.~Kossmann, ``On human intellect and
  machine failures: Troubleshooting integrative machine learning systems,'' in
  \emph{AAAI}, 2017.

\bibitem{alexander2016task}
E.~Alexander and M.~Gleicher, ``Task-driven comparison of topic models,''
  \emph{IEEE TVCG}, vol.~22, no.~1, 2016.

\bibitem{mcmahan2013ad}
H.~B. McMahan, G.~Holt, D.~Sculley, M.~Young, D.~Ebner, J.~Grady, L.~Nie,
  T.~Phillips, E.~Davydov, D.~Golovin, S.~Chikkerur, D.~Liu, M.~Wattenberg,
  A.~M. Hrafnkelsson, T.~Boulos, and J.~Kubica, ``Ad click prediction: A view
  from the trenches,'' in \emph{KDD}, 2013.

\bibitem{kahng2016visual}
M.~Kahng, D.~Fang, and D.~H.~P. Chau, ``Visual exploration of machine learning
  results using data cube analysis,'' in \emph{SIGMOD Workshop on
  Human-In-the-Loop Data Analytics}, 2016.

\bibitem{yu2014visualizing}
W.~Yu, K.~Yang, Y.~Bai, H.~Yao, and Y.~Rui, ``Visualizing and comparing
  convolutional neural networks,'' \emph{arXiv:1412.6631}, 2014.

\bibitem{wattenberg2016how}
M.~Wattenberg, F.~Viégas, and I.~Johnson, ``How to use {t-SNE} effectively,''
  \emph{Distill}, 2016.

\bibitem{maaten2008visualizing}
L.~v.~d. Maaten and G.~Hinton, ``Visualizing data using {t-SNE},'' \emph{JMLR},
  vol.~9, no. Nov, 2008.

\bibitem{nguyen2016multifaceted}
A.~Nguyen, J.~Yosinski, and J.~Clune, ``Multifaceted feature visualization:
  Uncovering the different types of features learned by each neuron in deep
  neural networks,'' in \emph{ICML Workshop on Vis for Deep Learning}, 2016.

\bibitem{rauber2016visualizing}
P.~E. Rauber, A.~X. Falc{\~a}o, and A.~C. Telea, ``Visualizing time-dependent
  data using dynamic t-sne,'' \emph{EuroVis}, vol.~2, no.~5, 2016.

\bibitem{liu2018visual}
S.~Liu, P.-T. Bremer, J.~J. Thiagarajan, V.~Srikumar, B.~Wang, Y.~Livnat, and
  V.~Pascucci, ``Visual exploration of semantic relationships in neural word
  embeddings,'' \emph{IEEE TVCG}, vol.~24, no.~1, 2018.

\bibitem{vondrick2013hoggles}
C.~Vondrick, A.~Khosla, T.~Malisiewicz, and A.~Torralba, ``Hoggles: Visualizing
  object detection features,'' in \emph{ICCV}, 2013.

\bibitem{rong2014word2vec}
X.~Rong, ``word2vec parameter learning explained,'' \emph{arXiv:1411.2738},
  2014.

\bibitem{park2018conceptvector}
D.~Park, S.~Kim, J.~Lee, J.~Choo, N.~Diakopoulos, and N.~Elmqvist,
  ``{ConceptVector}: text visual analytics via interactive lexicon building
  using word embedding,'' \emph{IEEE TVCG}, vol.~24, no.~1, 2018.

\bibitem{weld2018intelligible}
D.~S. Weld and G.~Bansal, ``Intelligible artificial intelligence,''
  \emph{arXiv:1803.04263}, 2018.

\bibitem{olah2017research}
C.~Olah and S.~Carter, ``Research debt,'' \emph{Distill}, 2017.

\bibitem{olah2017feature}
C.~Olah, A.~Mordvintsev, and L.~Schubert, ``Feature visualization,''
  \emph{Distill}, 2017.

\bibitem{mahendran2016visualizing}
A.~Mahendran and A.~Vedaldi, ``Visualizing deep convolutional neural networks
  using natural pre-images,'' \emph{IJCV}, vol. 120, no.~3, 2016.

\bibitem{grun2016taxonomy}
F.~Gr{\"u}n, C.~Rupprecht, N.~Navab, and F.~Tombari, ``A taxonomy and library
  for visualizing learned features in convolutional neural networks,''
  \emph{ICML Workshop on Vis for Deep Learning}, 2016.

\bibitem{kindermans2017patternnet}
P.-J. Kindermans, K.~T. Sch{\"u}tt, M.~Alber, K.-R. M{\"u}ller, and
  S.~D{\"a}hne, ``Learning how to explain neural networks: Patternnet and
  patternattribution,'' \emph{arXiv:1705.05598}, 2017.

\bibitem{li2017beyond}
H.~Li, K.~Mueller, and X.~Chen, ``Beyond saliency: understanding convolutional
  neural networks from saliency prediction on layer-wise relevance
  propagation,'' \emph{arXiv:1712.08268}, 2017.

\bibitem{smilkov2017smoothgrad}
D.~Smilkov, N.~Thorat, B.~Kim, F.~Vi{\'e}gas, and M.~Wattenberg,
  ``{SmoothGrad}: removing noise by adding noise,'' in \emph{ICML Workshop on
  Vis for Deep Learning}, 2017.

\bibitem{zhou2016learning}
B.~Zhou, A.~Khosla, A.~Lapedriza, A.~Oliva, and A.~Torralba, ``Learning deep
  features for discriminative localization,'' in \emph{CVPR}, 2016.

\bibitem{dosovitskiy2016inverting}
A.~Dosovitskiy and T.~Brox, ``Inverting visual representations with
  convolutional networks,'' in \emph{CVPR}, 2016.

\bibitem{mahendran2015understanding}
A.~Mahendran and A.~Vedaldi, ``Understanding deep image representations by
  inverting them,'' in \emph{CVPR}, 2015.

\bibitem{zhou2014object}
B.~Zhou, A.~Khosla, A.~Lapedriza, A.~Oliva, and A.~Torralba, ``Object detectors
  emerge in deep scene {CNNs},'' \emph{arXiv:1412.6856}, 2014.

\bibitem{mordvintsev2015inceptionism}
A.~Mordvintsev, C.~Olah, and M.~Tyka, ``Inceptionism: Going deeper into neural
  networks,'' \emph{Google Research Blog}, 2015.

\bibitem{springenberg2014striving}
J.~T. Springenberg, A.~Dosovitskiy, T.~Brox, and M.~Riedmiller, ``Striving for
  simplicity: The all convolutional net,'' \emph{arXiv:1412.6806}, 2014.

\bibitem{wei2015understanding}
D.~Wei, B.~Zhou, A.~Torrabla, and W.~Freeman, ``Understanding intra-class
  knowledge inside {CNN},'' \emph{arXiv:1507.02379}, 2015.

\bibitem{nguyen2016plug}
A.~Nguyen, J.~Yosinski, Y.~Bengio, A.~Dosovitskiy, and J.~Clune, ``Plug \& play
  generative networks: Conditional iterative generation of images in latent
  space,'' \emph{arXiv:1612.00005}, 2016.

\bibitem{bengio2009learning}
Y.~Bengio \emph{et~al.}, ``Learning deep architectures for ai,''
  \emph{Foundations and trends in Machine Learning}, vol.~2, no.~1, 2009.

\bibitem{carter2017using}
S.~Carter and M.~Nielsen, ``Using artificial intelligence to augment human
  intelligence,'' \emph{Distill}, 2017.

\bibitem{vinyals2015show}
O.~Vinyals, A.~Toshev, S.~Bengio, and D.~Erhan, ``Show and tell: A neural image
  caption generator,'' in \emph{CVPR}, 2015.

\bibitem{antol2015vqa}
S.~Antol, A.~Agrawal, J.~Lu, M.~Mitchell, D.~Batra, C.~Lawrence~Zitnick, and
  D.~Parikh, ``Vqa: Visual question answering,'' in \emph{ICCV}, 2015.

\bibitem{goodfellow2014generative}
I.~Goodfellow, J.~Pouget-Abadie, M.~Mirza, B.~Xu, D.~Warde-Farley, S.~Ozair,
  A.~Courville, and Y.~Bengio, ``Generative adversarial nets,'' in \emph{NIPS},
  2014.

\bibitem{goodfellow2016nips}
I.~Goodfellow, ``{NIPS} 2016 tutorial: Generative adversarial networks,''
  \emph{arXiv:1701.00160}, 2016.

\bibitem{liu2017visualizing}
S.~Liu, D.~Maljovec, B.~Wang, P.-T. Bremer, and V.~Pascucci, ``Visualizing
  high-dimensional data: Advances in the past decade,'' \emph{IEEE TVCG},
  vol.~23, no.~3, 2017.

\bibitem{samek2017evaluating}
W.~Samek, A.~Binder, G.~Montavon, S.~Lapuschkin, and K.-R. M{\"u}ller,
  ``Evaluating the visualization of what a deep neural network has learned,''
  \emph{IEEE transactions on neural networks and learning systems}, 2017.

\bibitem{ribeiro2016should}
M.~T. Ribeiro, S.~Singh, and C.~Guestrin, ``Why should {I} trust you?:
  Explaining the predictions of any classifier,'' in \emph{KDD}, 2016.

\bibitem{tsa2016icharacterizing}
C.-Y. Tsai and D.~D. Cox, ``Characterizing visual representations within
  convolutional neural networks: Toward a quantitative approach,'' \emph{ICML
  Workshop on Vis for Deep Learning}, 2016.

\bibitem{ritter2017cognitive}
S.~Ritter, D.~G. Barrett, A.~Santoro, and M.~M. Botvinick, ``Cognitive
  psychology for deep neural networks: A shape bias case study,''
  \emph{arXiv:1706.08606}, 2017.

\bibitem{das2017human}
A.~Das, H.~Agrawal, L.~Zitnick, D.~Parikh, and D.~Batra, ``Human attention in
  visual question answering: Do humans and deep networks look at the same
  regions?'' \emph{Computer Vision and Image Understanding}, 2017.

\bibitem{tam2017analysis}
G.~K. Tam, V.~Kothari, and M.~Chen, ``An analysis of machine-and
  human-analytics in classification,'' \emph{IEEE TVCG}, vol.~23, no.~1, 2017.

\bibitem{barocas2016big}
S.~Barocas and A.~D. Selbst, ``Big data's disparate impact,'' \emph{Calif. L.
  Rev.}, vol. 104, pp. 671--769, 2016.

\bibitem{caliskan2017semantics}
A.~Caliskan, J.~J. Bryson, and A.~Narayanan, ``Semantics derived automatically
  from language corpora contain human-like biases,'' \emph{Science}, vol. 356,
  no. 6334, 2017.

\bibitem{wattenberg2016attacking}
\BIBentryALTinterwordspacing
M.~Wattenberg, F.~Viegas, and M.~Hardt, ``Attacking discrimination with smarter
  machine learning,'' \emph{Google Research Website}, 2016. [Online].
  Available:
  \url{https://research.google.com/bigpicture/attacking-discrimination-in-ml/}
\BIBentrySTDinterwordspacing

\bibitem{2017facets}
\BIBentryALTinterwordspacing
``Facets,'' \emph{Google PAIR}, 2017. [Online]. Available:
  \url{https://pair-code.github.io/facets/}
\BIBentrySTDinterwordspacing

\bibitem{hardt2016equality}
M.~Hardt, E.~Price, N.~Srebro \emph{et~al.}, ``Equality of opportunity in
  supervised learning,'' in \emph{NIPS}, 2016.

\bibitem{wall2017warning}
E.~Wall, L.~Blaha, L.~Franklin, and A.~Endert, ``Warning, bias may occur: A
  proposed approach to detecting cognitive bias in interactive visual
  analytics,'' \emph{VAST}, 2017.

\bibitem{szegedy2013intriguing}
C.~Szegedy, W.~Zaremba, I.~Sutskever, J.~Bruna, D.~Erhan, I.~Goodfellow, and
  R.~Fergus, ``Intriguing properties of neural networks,''
  \emph{arXiv:1312.6199}, 2013.

\bibitem{goodfellow2014explaining}
I.~J. Goodfellow, J.~Shlens, and C.~Szegedy, ``Explaining and harnessing
  adversarial examples,'' \emph{ICLR}, 2014.

\bibitem{nguyen2015deep}
A.~Nguyen, J.~Yosinski, and J.~Clune, ``Deep neural networks are easily fooled:
  High confidence predictions for unrecognizable images,'' in \emph{CVPR},
  2015.

\bibitem{metzen2017detecting}
J.~H. Metzen, T.~Genewein, V.~Fischer, and B.~Bischoff, ``On detecting
  adversarial perturbations,'' in \emph{ICLR}, 2017.

\bibitem{gu2014towards}
S.~Gu and L.~Rigazio, ``Towards deep neural network architectures robust to
  adversarial examples,'' \emph{arXiv:1412.5068}, 2014.

\bibitem{papernot2016distillation}
N.~Papernot, P.~McDaniel, X.~Wu, S.~Jha, and A.~Swami, ``Distillation as a
  defense to adversarial perturbations against deep neural networks,'' in
  \emph{Security and Privacy}, 2016.

\bibitem{das2018shield}
N.~Das, M.~Shanbhogue, S.-T. Chen, F.~Hohman, S.~Li, L.~Chen, M.~E. Kounavis,
  and D.~H. Chau, ``Shield: Fast, practical defense and vaccination for deep
  learning using jpeg compression,'' \emph{arXiv:1802.06816}, 2018.

\bibitem{kurakin2016adversarial}
A.~Kurakin, I.~Goodfellow, and S.~Bengio, ``Adversarial examples in the
  physical world,'' \emph{arXiv:1607.02533}, 2016.

\end{thebibliography}

%
%

%

\begin{IEEEbiography}[{\includegraphics[width=1in,height=1.25in,clip,keepaspectratio]{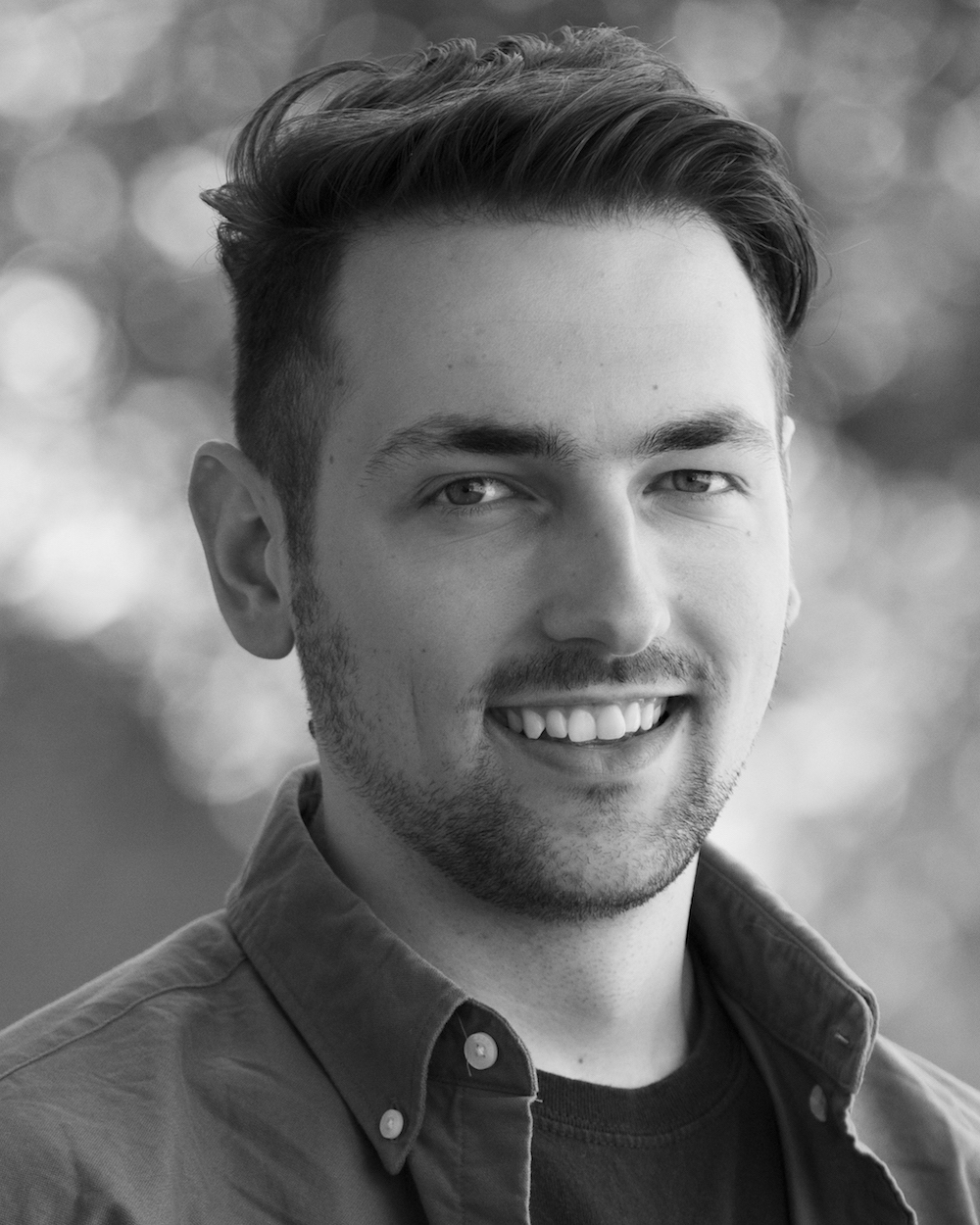}}]{Fred Hohman}
is a PhD student at Georgia Tech's College of Computing.
His research combines HCI principles and ML techniques to improve deep learning interpretability.
He won the NASA Space Technology Research Fellowship.
He received his B.S. in mathematics and physics.
He won SIGMOD'17 Best Demo, Honorable Mention;  Microsoft AI for Earth Award for using AI to improve sustainability; and the President's Fellowship for top incoming PhD students.
\end{IEEEbiography}

\begin{IEEEbiography}[{\includegraphics[width=1in,height=1.25in,clip,keepaspectratio]{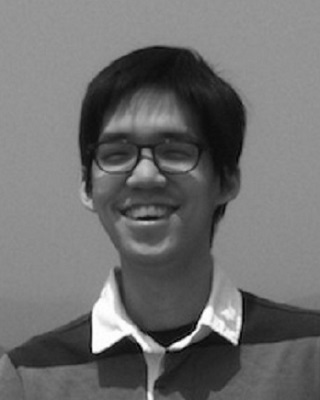}}]{Minsuk Kahng}
is a computer science PhD student at Georgia Tech.
His thesis research focuses on building visual analytics tools for exploring, interpreting, and interacting with complex machine learning models and results, by combining methods from information visualization, machine learning, and databases.
He received the Google PhD Fellowship and NSF Graduate Research Fellowship.
His ActiVis deep learning visualization system has been deployed on Facebook's machine learning platform.
\end{IEEEbiography}

\begin{IEEEbiography}[{\includegraphics[width=1in,height=1.25in,clip,keepaspectratio]{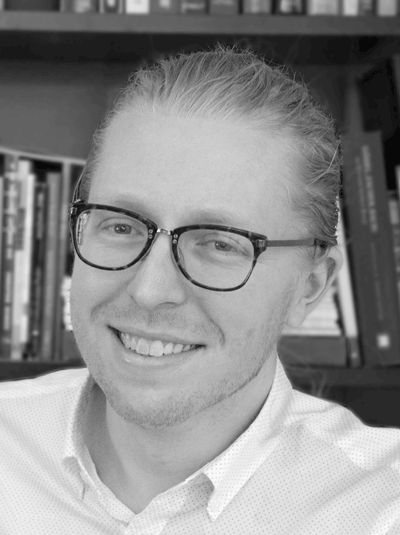}}]{Robert Pienta} is an industry researcher in applied machine learning and visual analytics. He received his PhD degree in computational science and engineering from Georgia Tech in 2017. He was an NSF FLAMEL fellow and presidential scholar at Georgia Tech.   His research interests include visual analytics, graph analytics, and machine learning. In particular, the algorithms and design techniques for interactive graph querying and exploration. 
\end{IEEEbiography}


\begin{IEEEbiography}[{\includegraphics[width=1in,height=1.25in,clip,keepaspectratio]{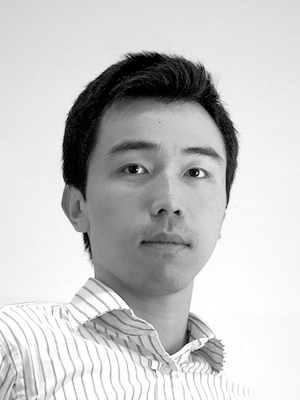}}]{Duen Horng (Polo) Chau} is an Associate Professor at Georgia Tech. 
His research bridges data mining and HCI to make sense of massive datasets.
His thesis won Carnegie Mellon's CS Dissertation Award, Honorable Mention. 
He received awards from Intel, Google, Yahoo, LexisNexis, and Symantec; 
He won paper awards at SIGMOD, KDD and SDM.
He is an ACM IUI steering committee member, IUI’15 co-chair, and IUI’19 program co-chair.
His research is deployed by Facebook, Symantec, and Yahoo. 
\end{IEEEbiography}




\end{document}